\documentclass[article]{JHEP3}
\usepackage{graphicx}
\usepackage{epsf,colordvi,shadow,pifont}
\usepackage{amsmath,epsfig}
\usepackage{amssymb,amsfonts}
\usepackage{latexsym}
\usepackage{epsfig}
\usepackage[vcentermath,enableskew]{youngtab}
\def\hri#1#2{\href{http://arxiv.org/abs/#1}{[ArXiv:#1]#2}}
\def\hre#1#2{\href{http://arxiv.org/abs/#1/#2}{[ArXiv:#1/#2]}}
\def\hspi#1#2{\href{http://www.slac.stanford.edu/spires/find/hep/www?irn=#1}{#2}}
\relax

\def\be{\begin{equation}}
\def\ee{\end{equation}}

\newcommand{\bear}{\begin{eqnarray}}
\newcommand{\bea}{\begin{eqnarray}}
\newcommand{\eear}{\end{eqnarray}}
\newcommand{\eea}{\end{eqnarray}}
\def\hre#1#2{\href{http://arxiv.org/abs/#1/#2}{[ArXiv:#1/#2]}}
\def\hspi#1#2{\href{http://www.slac.stanford.edu/spires/find/hep/www?irn=#1}{#2}}
\newbox\pippobox

\renewcommand{\b}[1]{\textbf{#1}}

\def\II{\relax{\rm I\kern-.18em I}}

\def\m{\mu}

\def\s{\sigma}

\def\sp{\;\;\;,\;\;\;}

\def\f{\varphi}
\def\z{\zeta}
\def\a{\alpha}

\def\tr{\ensuremath{\mathrm{Tr}}}

\def\l{\lambda}
\def\nn{\nonumber}
\def\b{\beta}
\newcommand{\<}{\langle}
\renewcommand{\>}{\rangle}
\newbox\pippobox

\newcommand{\intd}{\mathrm{d}}

\def\ls{\ell_s}

\def\lab{\label}
\newcommand{\de}{\partial}
\def\le{\left}
\def\ri{\right}

\title{Improved Holographic QCD}

\author{Umut G\"ursoy$^1$, Elias Kiritsis$^{2}$\footnote{On leave of absence from APC,
Universit\'e Paris 7, (UMR du CNRS 7164).}, Liuba Mazzanti$^3$, Georgios Michalogiorgakis$^4$ and Francesco Nitti$^5$\\

$^1$\href{http://www1.phys.uu.nl/wwwitf}{Institute for Theoretical Physics, Utrecht University;
Leuvenlaan 4, 3584 CE Utrecht, The Netherlands.}\\
~\\
$^2$\href{http://hep.physics.uoc.gr/}{ Crete Center for Theoretical Physics, Department of Physics, University of Crete
71003 Heraklion, Greece}\\
~\\
$^3$\href{http://www-fp.usc.es/~theory/}{Departamento de F\'{\i}sica de Part\'{\i}culas, Universidade
de Santiago de
Compostela\\and\\Instituto Galego de F\'{\i}sica de Altas
Enerx\'{\i}as (IGFAE)\\E-15782, Santiago de Compostela, Spain}\\
~\\
$^4$Department of Physics, Purdue University, 525 Northwestern Avenue, West Lafayette, IN 47907-2036\\
~\\
$^5$\href{http://www.apc.univ-paris7.fr}{APC, Universit\'e Paris 7, \\ B\^atiment Condorcet, F-75205, Paris Cedex 13, France\\
 (UMR du CNRS 7164).}}
\preprint{\phantom{\hepth{yymmyyyy}}\\CCTP-2010-7}

\abstract{We provide a review to holographic models based on Einstein-dilaton
 gravity with a potential in 5 dimensions. Such theories, for a judicious choice of potential
are very close to the physics of large-N YM theory both at zero and finite
 temperature. The zero temperature glueball spectra as well as their finite temperature thermodynamic functions
 compare well with lattice data. The model can be used to calculate transport coefficients, like bulk viscosity, the drag force and jet quenching parameters,
 relevant for the physics of the Quark-Gluon Plasma.     }

\keywords{Gauge-gravity correspondence,  QCD, Quark Gluon Plasma, Holographic Gauge theory\\
~\\
Based on lectures given at the 5th Aegean Summer School (Milos, Greece), September 2009. To appear in the proceedings}

\begin{document}

\section{Introduction}
The experimental efforts at RHIC, \cite{kirrhic} have provided a novel window in the physics of the strong interactions.
The consensus on the existing data is that shortly after the collision, a ball of quark-gluon plasma (QGP) forms that is at thermal equilibrium,
and  subsequently expands until its temperature falls below the QCD transition (or crossover) where it finally hadronizes.
Relativistic hydrodynamics describes very well the QGP \index{Quark Gluon Plasma} \cite{kirlr},
with a shear-viscosity to entropy density ratio close to that
\index{viscosity!shear}
of ${\cal N}=4$ SYM, \cite{kirpss}.
The QGP is at strong coupling, and it necessitates a treatment beyond perturbative QCD approaches, \cite{kirreview}.
Moreover, although the shear viscosity from  ${\cal N}=4$ seems to be close to that ``measured'' by
 experiment, lattice data indicate that in the relevant
RHIC range  $1\leq {T\over T_c}\leq 3$ the QGP
seems not to be a fully  conformal fluid.
 Therefore the bulk viscosity \index{viscosity!bulk} may play a role near the phase transition
\cite{kirkkt,kirm}.  The lattice techniques have been successfully used
to study the thermal behavior of QCD, however they are not easily
extended to the computation of hydrodynamic quantities. They can
be used however, together with parametrizations of correlators in
order to pin down parameters \cite{kirm}. On the other hand,
approaches based on holography have the potential to address
directly the real-time strong coupling physics relevant for
experiment.

In the bottom-up holographic model of AdS/QCD \cite{kiradsqcd1}, the bulk viscosity is zero as conformal invariance is essentially not broken
(the stress tensor is traceless).
In the soft-wall model \cite{kirsoft}, \index{soft wall model} no reliable calculation can be done for glue correlators and therefore
transport coefficients are ill-defined. Similar
remarks hold for other phenomenologically interesting observables as the
drag force \index{drag force}  and the jet quenching parameter \index{jet quenching} \cite{kirher,kirlrw,kirgub,kirtea}.

Top-down holographic models of QCD displaying all relevant features of the theory have been difficult to obtain.
Bottom-up models based on AdS slices \cite{kirps} have given some insights mostly in the meson sector, \cite{kiradsqcd1}
but necessarily lack many important holographic features of QCD.
A hybrid approach has been advocated \cite{kirihqcd,kirdiss} combining features of bottom-up and top-down models.
An similar approach was proposed independently in \cite{kirgubser}.
Such an approach, called Improved Holographic QCD (or IHQCD for short)
\index{Improved Holographic QCD}
is essentially a five-dimensional dilaton-gravity system with a non-trivial dilaton potential.
Flavor can be eventually added in the form of $N_f$ space-time filling
$D4-\overline{D4}$ brane pairs, supporting $U(N_f)_L\times U(N_f)_R$ gauge fields and a bi-fundamental
scalar \cite{kirckp}.
The UV asymptotics of the potential are fixed by QCD perturbation theory,
while the IR asymptotics of the potential can be fixed by confinement and linear glueball asymptotics.
An analysis of the finite temperature behavior \cite{kirGKMN1,kirGKMN2} has shown that the phase structure is exactly what one would expect from YM.
A potential with a single free parameter tuned to match
 the zero temperature glueball spectrum was able to agree with the thermodynamic behavior of glue
to a good degree, \cite{kirGKMN1}. Similar results, but with somewhat different potentials were also obtained in \cite{kirgubser,kirdew}

In \cite{kirGKMN1,kirGKMN2,kirgubser} it was shown that Einstein-dilaton gravity \index{Einstein-dilaton gravity} with a strictly monotonic dilaton potential that grows sufficiently
fast, generically shares the same phase structure and
thermodynamics of finite-temperature pure Yang-Mills theory at large $N_c$.
There is a deconfinement phase transition (dual to a Hawking-Page phase transition
between a black hole\index{black hole} and thermal gas\index{thermal gas}
background on the gravity side), which is generically first order.
 The latent heat scales as $N_c^2$. In the deconfined gluon-plasma phase,
the free energy slowly approaches that of a free gluon gas at high temperature,
and the speed of sound starts from a small value at $T_c$ and approaches the conformal value $c_s^2=1/3$ as
 the temperature increases. The deviation from conformal invariance is strongest at $T_c$, and
is signaled by the presence of a non-trivial gluon condensate, \index{gluon condensate} which on the gravity side emerges as
a deviation of the scalar solution that behaves asymptotically as $r^4$ close to the UV boundary. In the CP-violating
sector, the topological vacuum density \index{topological vacuum density} $\tr F\tilde{F}$ has zero expectation value in the deconfined phase,
in agreement with lattice results \cite{kirltheta} and large-$N_c$ expectations.

The analysis performed in \cite{kirGKMN2} was completely general and did not rely on any specific form of the dilaton potential $V(\l)$.
A  detailed  analysis of an explicit model in \cite{kirfit} shows that the thermodynamics matches {\em quantitatively}
the thermodynamics of pure Yang-Mills theory.
The (dimensionless) free energy, entropy density, latent heat and speed of sound, obtained on the gravity side by numerical integration of the 5D
field equations, can be compared with the corresponding quantities, calculated on the lattice  for pure Yang-Mills at finite-$T$, resulting in excellent
agreement, for the temperature range that is accessible by lattice techniques.
The same  model also shows a good agreement with the lattice calculation of glueball mass ratios \index{glueball mass ratios} at zero temperature,
and  the  value of the deconfining
critical temperature (in units of the lowest glueball mass)  is also in good agreement with the lattice results.

In short, the model we present gives a good phenomenological holographic description of
most static  properties\footnote{There are very few observables also that are not in agreement with YM.
 They are discussed in detail in \cite{kirdiss}.} (spectrum and equilibrium thermodynamics)  of
large-$N_c$ pure Yang-Mills, as computed on the lattice, for energies up to several  times $T_c$.  Thus it constitutes a good starting point for the
computation of dynamical observables in a {\em realistic} holographic dual to QCD (as opposed to e.g. ${\cal N}=4$ SYM), such as transport
coefficients and other hydrodynamic properties that are not easily accessible by lattice techniques, at energies and temperatures relevant for
relativistic heavy-ion collision experiments. We will report on such a calculation in the near future.

The vacuum solution in this model is described in terms of two basic bulk fields, the metric and the dilaton.
These are not the only bulk fields however, as the bulk theory is expected to have an a priori infinite number
of fields, dual to all possible YM operators.
In particular we know from the string theory side that there are a few other low mass fields, namely the RR axion \index{RR axion}
(dual to the QCD $\theta$-angle)\index{$\theta$-angle}
the NSNS and RR two forms $B_2$ and $C_2$ as well as other higher-level fields.
 With the exception of the RR axion, such fields are dual
to higher-dimension and/or higher-spin operators of YM.
Again, with the exception of the RR axion, they are not expected to play an important
 role into the structure of the vacuum and this is why we neglect them when we solve the equations
of motion. However, they are going to generate several new towers of glueball states
 beyond those that we discuss in this paper (namely the $0^{++}$
glueballs associated to dilaton fluctuations, $2^{++}$
glueballs associated to graviton  fluctuations and $0^{-+}$
glueballs associated to RR axion  fluctuations). Such fields can be included in
 the effective action and the associated glueball spectra calculated.
Since we do not know the detailed structure of the associated string theory,
their effective action will depend on more semi-phenomenological functions like $Z(\lambda)$ in (\ref{kiraxionaction}).
These functions can again be determined in a way similar to $Z(\lambda)$.
In particular including the $B_2$ and $C_2$ field will provide $1^{+-}$ glueballs among others.
Fields with spin greater than 2 are necessarily stringy in origin.
We will not deal further with extra fields, like $B_2$ and $C_2$ and other as they are not particularly relevant for the purposes of
this model, namely the study of finite temperature physics in the deconfined case. We will only consider the axion,
as its physics is related to the CP-odd sector of YM with an obvious phenomenological importance.

It is well documented that string theory duals of YM must have strong curvatures in the UV regime.
This has been explained in detail in \cite{kirdiss}
where it was also argued, that although the asymptotic AdS boundary
geometry is due to the curvature non-linearities of the associated string theory,
the inwards geometry is perturbative around AdS, with logarithmic corrections, generating the YM perturbation theory.
The present model is constructed so that it takes the asymptotic AdS geometry for granted, by introducing the associated vacuum energy by hand,
and simulates the perturbative YM expansion by an appropriate dilaton potential.
In the IR, we do not expect strong curvatures in the string frame, and indeed the preferred backgrounds have this property.
In this sense the model contains in itself the relevant expected effects that should arise from strong curvatures in all regimes.
These issues have been explained in \cite{kirihqcd} and in more detail in \cite{kirdiss}.

A different and interesting direction is the use of such models to study the expansion of the plasma and the associated dynamics.
Such a context is similar to what happens on cosmology, especially the one related to the Randall-Sundrum setup.
Indeed in this case the expansion can be found by following the geodesic motion of probe
branes in the relevant background, \cite{kircosmo0}-\cite{kircosmo3}.\index{holographic cosmology}
This generalizes to more complicated backgrounds, \cite{kircosmo3}, like the ones studied here.

Once we have a holographic model we trust, we should calculate observables ,
 like transport coefficients\index{transport coefficients}  that are hard to calculate on the lattice.
A first class of transport coefficients are viscosity \index{viscosity} coefficients.\footnote{These
are the leading transport coefficients in the derivative expansion.
There are subleading coefficients that have been calculated recently for ${\cal N}=4$ SYM,
 \cite{kirsub}. However, at the present level of accuracy,
 they cannot affect substantially the comparison to experimental data, \cite{kirlr}.}
A general fluid is characterized by two viscosity coefficients, the shear $\eta$ \index{viscosity|shear}
and the bulk viscosity $\zeta$.\index{viscosity!bulk}
The shear viscosity in strongly coupled theories described by gravity duals was shown to be universal, \cite{kirpss}.
In particular, the ratio $\eta/s$, with $s$ the entropy density, is equal to ${1\over 4\pi}$.
This is correlated to the universality of low-energy scattering of gravitons from black holes.
It is also known that deviations from this value can only be generated by higher curvature terms that contain the Riemann tensor
(as opposed to the Ricci tensor of the scalar curvature).
In QCD, as the theory is strongly coupled in the temperature range $T_c\leq T\leq 3T_c$,
we would expect that $\eta/s\simeq {1\over 4\pi}$.
Recent lattice calculations, \cite{kirmshear} agree with this expectations although potential systematic
 errors in lattice calculations of transport coefficients can be large.

Conformal invariance forces the bulk viscosity to vanish. Therefore the  ${\cal N}=4$ SYM plasma, being a conformal fluid, has vanishing bulk viscosity.
QCD on the other hand is not a conformal theory. The classical theory is however conformally invariant and asymptotic freedom implies that
conformal invariance is a good approximation in the UV.
 This would suggest that the bulk viscosity to entropy ratio  is negligible at large temperatures.
   However it is not expected to be so in the IR: as mentioned earlier lattice data indicate that in the relevant
RHIC range  $1\leq {T\over T_c}\leq 3$ the QGP
seems not to be a fully  conformal fluid.
 Therefore the bulk viscosity may play a role near the phase transition.

 So far there have been two approaches that have calculated the bulk viscosity\index{viscosity!bulk}
 in YM/QCD, \cite{kirViscos1,kirViscos2,kirsum,kirViscos3} and have both indicated that the bulk viscosity
 rises near the phase transition as naive expectation would suggest.
 The first used the method of sum rules in conjunction with input from Lattice thermodynamics, \cite{kirViscos1,kirViscos2,kirsum}.
 It suggested a dramatic rise of the bulk viscosity near $T_c$ although the absolute normalization of the result is uncertain.
 The reason is that this method relies on an ansatz for the density associated with stress-tensor two point functions that are otherwise unknown.

 The second method \cite{kirViscos3} relies on a direct computation of the density at low frequency of the appropriate stress-tensor two-point function.
 As this computation is necessarily Euclidean, an analytic continuation is necessary.  The values at a finite number of discrete Matsubara frequencies
 are not enough to analytically continue. An ansatz for the continuous density is also used here, which presents again a potentially
 large systematic uncertainty.

Calculations in IHQCD support a rise of the bulk viscosity near $T_c$, but the values are much smaller than previously expected.
 Studies of how this affects hydrodynamics at RHIC, \cite{kirheinz} suggest that this implies a  fall in radial and elliptic flow.

Another class of interesting experimental observables is associated with quarks, and comes under
the label of ``jet quenching". \index{jet quenching} Central to this is the expectation that an energetic quark will loose energy very fast in the quark-gluon plasma
because of strong coupling. This has as a side effect that back-to back jets are suppressed. Moreover if a pair of energetic quarks is generated
near the plasma boundary then one will exit fast the plasma and register as an energetic jet, while the other will thermalize and its
identity will disappear. This has been clearly observed at RHIC and used to study the energy loss of quarks in the quark-gluon plasma.

Heavy quarks\index{heavy quarks}  are of extra importance, as their mass masks some low-energy
 strong interaction effects, and can be therefore cleaner probes of plasma energy loss.
There are important electron observables at RHIC, \cite{kirphenix} that can probe heavy-quark energy loss in the strongly coupled quark-gluon plasma.
Such observables are also expected to play an important role in LHC \cite{kirurs}.\index{RHIC}\index{LHC}

A perturbative QCD approach to calculate the energy loss of a heavy quark in the plasma has been pursued by calculating radiative energy loss, \cite{kirrel}.
However its application to the RHIC plasma has recently raised  problems,
 based on comparison with data. A phenomenological coefficient used in such cases is known as the jet quenching coefficient
$\hat q$, and is defined as the rate of change of the average value of transverse momentum square of a probe.
Current fits,  \cite{kirphenix,kirlan}, indicate that a value of order 10 $GeV^2/fm$ or more is needed to describe the data
 while perturbative approaches are trustworthy at much lower values.

Several attempts were made to compute quark energy loss in the holographic context, relevant for   ${\cal N}=4$ SYM\footnote{Most are reviewed in
\cite{kirgubser-review}.}. In some of them \cite{kirLiu1,kirLiu2} the jet-quenching coefficient $\hat q$ was
calculated via its relationship to a light-like Wilson loop.
Holography was then used to calculate the appropriate Wilson loop.
   The $\hat q$ obtained scales as $\sqrt{\lambda}$ and as the third power of the temperature,
   \be
\hat q_{\rm conformal}={\Gamma\left[{3\over 4}\right]\over \Gamma\left[{5\over 4}\right]}~\sqrt{2\l}~\pi^{3\over 2}T^3
\ee

A different approach chooses to compute the drag force \index{drag force} acting a string whose UV end-point
 (representing an infinitely heavy quark) is forced to move with constant velocity $v$, \cite{kirher,kirgub,kirtea}, in the context of
 ${\cal N}=4$ SYM plasma.
 The result for the drag force is
 \be
F_{\rm conformal}={\pi\over 2}\sqrt{\lambda}~T^2{v\over \sqrt{1-v^2}}\ee
and is calculated by first studying the equilibrium configuration of the appropriate string world-sheet string
and then calculating the momentum flowing down the string.
This can be the starting point of a Langevin evolution system,\index{Langevin diffusion}  as the process of
energy loss has a stochastic character, as was first pointed out in \cite{kirlan1} and more recently pursued in \cite{kirlan2}-\cite{kiriancu}.

Such a system involves a classical force, that in this case is the drag force, and a stochastic noise that is taken to be Gaussian
and which is characterized by a diffusion coefficient.
There are two ingredients here that are novel. The first  is that the Langevin evolution must be relativistic, as
the quarks can be very energetic. Such relativistic systems have been described in the mathematical physics literature, \cite{kirmath}
and have been used in phenomenological analyses of heavy-ion data, \cite{kirlan}.
They are known however to have peculiar behavior, since demanding an equilibrium relativistic Boltzmann
 distribution, provides an Einstein relation that is pathological
at large temperatures. Second, the transverse and longitudinal
diffusion coefficients are not the same, \cite{kirGubser-lan}. A
first derivation of such Langevin dynamics from holography was
given in \cite{kirGubser-lan}. This has been extended in
\cite{kiriancu} where the thermal-like noise  was associated and interpreted in
terms of the world-sheet horizon that develops on the probe
string.

Most of the transport properties mentioned above have been successfully computed in ${\cal N}=4$ SYM and a lot of
debate is still waged as to how they can be applied to QCD in the appropriate temperature range, \cite{kirgub},\cite{kircaron},\cite{kirsin}.
A holographic description of QCD has been elusive,  and the best we have so far have been simple bottom up models.

In the simplest bottom-up holographic model known as  AdS/QCD \cite{kiradsqcd1},
the bulk viscosity is zero as conformal invariance is essentially not broken
(the stress tensor is traceless),  and the drag force and jet quenching essentially retain their conformal values.

In the soft-wall model \cite{kirsoft}, no reliable calculation can be done for glue correlators and therefore
transport coefficients are ill-defined, as bulk equations of motion are not respected. Similar
remarks hold for other phenomenologically interesting observables as the
drag force and the jet quenching parameter.

The shear viscosity of IHQCD is the same as that of ${\cal N}=4$
\index{viscosity!shear}
SYM, as the model is a two derivative model. Although this is not
a good approximation in the UV of QCD, it is expected to be a good
approximation in the energy range $T_c\leq T\leq 5T_c$.
The bulk viscosity in IHQCD rises near the phase transition
 but ultimately stays slightly below the shear viscosity. There is a
  general holographic argument that any (large-N) gauge theory that confines color at zero temperature should
 have an increase in the bulk viscosity-to-entropy density ratio close to
 $T_c$.

The drag force on heavy quarks,  and the associated diffusion times,
can be calculated and found to be momentum depended as anticipated from asymptotic freedom.
Numerical values of diffusion times are in the region dictated by phenomenological analysis of heavy-ion data.
The medium-induced corrections to the  quark mass (needed
for the diffusion time calculation) can be calculated,
and they result in a mildly decreasing  effective quark  mass
as a function of temperature. This is consistent with lattice results.
Finally, the jet-quenching parameter can be calculated and found to be comparable at $T_c$
to the one obtained by extrapolation from
${\cal N}=4$ SYM. Its temperature dependence is however different
and again reflects the effects of asymptotic freedom.

\section{The 5D model} \label{kirmodel}

The holographic dual of large $N_c$ Yang Mills theory,  proposed in \cite{kirihqcd},  is
based on a five-dimensional Einstein-dilaton model, \index{Einstein-dilaton gravity} with the action\footnote{Similar models of Einstein-dilaton gravity were proposed
 independently in \cite{kirgubser} to describe the finite temperature physics of
  large $N_c$ YM. They differ in the UV as the dilaton corresponds to a relevant operator
 instead of the marginal case we study here. The gauge coupling $e^{\Phi}$
  also asymptotes to a constant instead of zero in such models.}:
\begin{equation}
S_5=-M^3_pN_c^2\int d^5x\sqrt{g}
\left[R-{4\over 3}(\partial\Phi)^2+V(\Phi) \right]+2M^3_pN_c^2\int_{\partial M}d^4x \sqrt{h}~K.
 \label{kira1}\end{equation}
Here, $M_p$ is the  five-dimensional Planck scale and $N_c$ is the number of colors.
The last term is the Gibbons-Hawking term, \index{Gibbons-Hawking term} with $K$ being the extrinsic curvature
of the boundary. The effective five-dimensional Newton constant
is $G_5 = 1/(16\pi M_p^3 N_c^2)$, and it is small in the large-$N_c$ limit.

Of the 5D coordinates $\{x_i, r\}_{i=0\ldots 3}$, $x_i$ are identified with the
4D space-time coordinates, whereas  the  radial coordinate $r$ roughly corresponds to the 4D RG scale.
We identify $\l\equiv e^\Phi$ with the  running 't Hooft  coupling $\l_t\equiv N_cg_{YM}^2$,
up to an {\it a priori} unknown multiplicative factor\footnote{ This relation is well motivated
in the UV, although it may be modified at strong coupling (see Section \ref{kirscheme}). The
quantities we will calculate do not depend on the explicit relation between $\l$ and $\l_t$.
}, $\l = \kappa \l_t$.

The dynamics is encoded in the dilaton potential\footnote{With a slight abuse of notation we will denote $V(\l)$  the
function $V(\Phi)$ expressed as a function of  $\l\equiv e^\Phi$.},  $V(\l)$.
The small-$\l$ and large-$\l$ asymptotics of $V(\l)$ determine the solution in the UV
and  the IR of the geometry
respectively. For a detailed but concise description of the UV and IR  properties of the solutions the reader
is referred to Section 2 of \cite{kirGKMN2}. Here we will only mention the most relevant information:
\begin{enumerate}
\item For small $\l$,   $V(\l)$  is required to have a power-law expansion of the form:
\be \label{kirUVexp}
V(\l) \sim {12\over \ell^2}(1+ v_0 \l + v_1 \l^2 +\ldots), \qquad \l\to 0 \;.
\ee
The value at $\l=0$ is constrained to be finite and positive, and sets the UV AdS scale $\ell$.
 The coefficients
of the other terms in the expansion fix the  $\beta$-function coefficients for the
running coupling $\l(E)$. If we  identify the energy scale with the metric scale factor in the Einstein frame, as
in  \cite{kirihqcd}, we obtain:
\bea\label{kirbetafunc}
&&\beta(\l) \equiv {d \lambda \over d\log E} = -b_0\l^2 -b_1 \l^3 +\ldots\nn\\
&& b_0 = {9\over 8} v_0, \quad \; \; b_1 = \frac94 v_1 - \frac{207}{256}v_0^2 \;.
\eea
\item For large $\l$,   confinement and the absence of \index{bad singularity} bad singularities\footnote{We call ``bad  singularities'' those that
do not have  a well defined spectral problem for the fluctuations
without imposing extra boundary conditions.} require:
\be\label{kirIRexp}
 V(\l) \sim \l^{2Q}(\log \l)^P \quad \l\to \infty, \quad \left\{ \begin{array}{l} 2/3 < Q < 2\sqrt{2}/3, \quad P\; {\rm arbitrary}\\ Q = 2/3, \quad P\geq 0 \end{array} \right. .
\ee
In particular, the values $Q=2/3, P=1/2$ reproduce an asymptotically-linear glueball spectrum, \index{glueball spectrum!linear}
$m_n^2\sim n$, besides confinement.
We will restrict ourselves to this case in what follows.
\end{enumerate}

In the large $N_c$ limit,
the  canonical ensemble partition function of the model just described, can be
approximated by a sum over saddle points, each given by a classical solution of the Einstein-dilaton
field equations:
\be
{\cal Z}(\beta) \simeq e^{-{\cal S}_1(\beta)}  +   e^{-{\cal S}_2(\beta)} + \ldots
\ee
where ${\cal S}_i$ are the euclidean actions evaluated on each  classical solution with a fixed
 temperature $T=1/\beta$, i.e. with euclidean time compactified on a circle of length $\beta$.
There are two possible types of Euclidean solutions which preserve 3-dimensional rotational invariance.
In conformal coordinates these are:
\begin{enumerate}
\item {\bf Thermal gas solution,}
\be\label{kirthermal}
ds^2 = b^2_o(r)\left(dr^2 + dt^2 +  dx_mdx^m\right), \qquad \Phi = \Phi_o(r),
\ee
with $r\in (0, \infty)$ for the values of $P$ and $Q$ we are using;
\item {\bf Black-hole solutions,}\index{black hole}
\be
ds^2=b(r)^2\left[{dr^2\over f(r)}+f(r)dt^2+dx_mdx^m\right], \qquad \Phi = \Phi(r),
 \label{kira7}\ee
with  $r\in (0,r_h)$,  such that $f(0)=1$, and $f(r_h)=0$.
\end{enumerate}
In both cases Euclidean time is periodic with period $\b_o$ and $\b$ respectively
 for the thermal gas and black-hole solution,
and 3-space is
taken to be a torus with volume $V_{3o}$ and $V_3$ respectively,
so that the black-hole mass and entropy are finite\footnote{The periods
and 3-space volumes of the thermal gas solution are related to the black-hole
solution values by requiring that the geometry of the two solutions are
the same on the (regulated) boundary. See \cite{kirGKMN2} for details.}.

The black holes are dual to a deconfined phase, since the string
tension vanishes at the horizon, and the Polyakov loop has\index{Polyakov loop}
non-vanishing expectation value (\cite{kirD4,kirsonnenschein}). On the
other hand, the thermal  gas \index{thermal gas} background is confining.

The thermodynamics of the deconfined phase\index{deconfined phase} is dual to the  5D
black-hole thermodynamics. The  free energy, defined as\index{black hole}
 \be\label{kirfirst law}
{\cal F} = E - T S,
\ee
is identified with the  black-hole on-shell
action; as usual, the energy $E$ and entropy $S$ are identified  with the black-hole mass, and one
fourth of the horizon area in Planck units,  respectively.

The thermal gas and black-hole solutions with the same temperature differ at $O(r^4)$:
\be
b(r) = b_o(r)\left[1 + \,{\cal G}\, {r^4\over \ell^3} +\ldots\right],  \qquad f(r) = 1 -{C\over 4} {r^4\over \ell^3} + \ldots \qquad r \to 0,
\label{kirb-bo}
\ee where ${\cal G}$ and $C$ are constants with units of energy.
As shown in \cite{kirGKMN2} they  are related to the enthalpy $TS$ and
the gluon condensate $\<\tr F^2\>$ : \be\label{kirCG} C = {T S \over
M_p^3 N_c^2 V_3} , \qquad \qquad {\cal G} ={22 \over 3 (4\pi)^{2}}
{\langle \tr~F^2 \rangle_T - \langle \tr~F^2 \rangle_o \over240
M_p^3 N_c^2}. \ee Although they appear as coefficients in the UV
expansion, $C$ and ${\cal G}$ are determined by regularity at the
black-hole horizon. For $T$ and $S$ the relation is the usual one,
\be\label{kirTS}
 T = - {\dot{f}(r_h) \over 4\pi}, \qquad S  = {Area \over 4 G_5} = 4\pi\, (M_p^3 N_c^2 V_3) \, b^3(r_h).
\ee
For ${\cal G}$ the relation with the horizon quantities is more complicated and
cannot be put in a simple analytic form. However, as discussed in \cite{kirGKMN2}, for each temperature
there exist only specific values of ${\cal G}$ (each corresponding to a different black hole)
such that the horizon is regular.

At any given temperature there can be one or more solutions: the thermal gas
is always present, and there can be different black holes\index{black hole} with the same temperature. The solution  that dominates
the partition function at a certain $T$ is the one with smallest free energy. The free energy difference between the black hole  and
thermal gas\index{thermal gas} was calculated in \cite{kirGKMN2}  to be:
\be\label{kirF}
\frac{\cal F}{M_p^3 N_c^2 V_3}={{\cal F}_{BH} - {\cal F}_{th}\over M_p^3 N_c^2 V_3}  = 15 {\cal G} - {C\over 4}.
\ee
For a dilaton potential corresponding to a confining theory, like the one we will assume,
the phase structure is the following \cite{kirGKMN2}:
\begin{enumerate}
\item There exists a minimum temperature $T_{min}$ below which the only solution is the thermal gas.
\item Two branches of black holes (``large'' and ``small'')  appear for $T\geq T_{min}$, but the ensemble
is still dominated by the confined phase up to a temperature $T_c > T_{min}$\index{critical temperature}
\item At $T=T_c$ there is a first order phase transition\index{phase transition}
to the {\em large} black-hole phase\index{black hole!large}. The system remains
in the black-hole (deconfined) phase for all $T>T_c$.
\end{enumerate}
In principle there could be more than two black-hole branches, but this
will not happen with the specific potential we will use.

\section{Scheme dependence}\label{kirscheme}

There are several sources of scheme dependence\index{scheme dependence} in any attempt to solve a QFT.
Different parametrizations of the coupling constant (here $\l$) give different descriptions.
However,  physical statements must be invariant under such a change.
In our case,  reparametrizations of the coupling constant are equivalent to radial diffeomorphisms as we could use
$\l$ as the radial coordinate.

In the holographic context,  scheme dependence related to
coupling redefinitions translates into field redefinitions for the bulk fields.
As the bulk theory is on-shell, all on-shell observables
(that are evaluated at the single boundary of space-time) are independent
of the field redefinitions showing that scheme-independence is expected.
Invariance under radial reparametrizations of scalar bulk invariants is equivalent to RG invariance.
Because of renormalization effects, the boundary is typically shifted and in this case field redefinitions must be combined
with appropriate radial diffeomorphisms that amount to RG-transformations.

Another source of scheme dependence in our setup comes from the choice of
 the energy function. Again we may also consider this as a radial coordinate and therefore
it is subject to coordinate transformations.
A relation between $\l$ and $E$ is the $\beta$-function\index{$\beta$-function},
\be
{d\l\over d\log E}=\beta(\l).
\ee
$\beta$ by definition transforms as a vector under $\l$ reparametrizations
and as a form under $E$ reparametrizations.
$\beta(\l)$ can therefore be thought of as a vector field implementing the change of coordinates from $\lambda$ to $E$ and vice-versa.

Physical quantities should be independent of scheme. They are
quantities that are fully diffeomorphism invariant. If the
gravitational theory had no boundary there would be no
diffeomorphism invariant quantities, except for possible
topological invariants. Since we have a boundary, diffeomorphism
invariant quantities are defined at the boundary.

Note that scalar quantities are not invariant. To be invariant they must be scalar and constant.
We therefore need to construct scalar functions that are invariant under changes of radial coordinates.

We can fix this reparametrization invariance by picking a very special frame.
For example choosing the (string) metric in the conformal frame
\be
ds^2=e^{2A}\left[dr^2+dx^{\m}dx_{\mu}\right]\sp \l(r)
\ee
or in the domain-wall frame
\be
ds^2=du^2+e^{2A}dx^{\m}dx_{\m}\sp \l(u)
\ee
fixes the radial reparametrizations almost completely. In conformal frame,
 common scalings of $r,x^{\mu}$ are allowed, corresponding to constant
shifts of $A(r)$.

Eventually we are led to calculate and compare our results to
other ways of calculating (like the lattice).
Some outputs are easier to compare (for example correlators).
Others are much harder as they are not invariant (like the value
 of the coupling at a given energy scale).

In the UV such questions are well understood. The asymptotic energy scale is fixed by comparison to conformal field theory
examples. This is possible because the space is asymptotically AdS$_5$\footnote{As
the dilaton is now not constant there is a non-trivial question: in which frame
is the metric
AdS. In \cite{kirdiss} it was argued that this should be the case in the string
frame. The difference of course between the string and Einstein frame is
 subleading  in the UV as the coupling constant vanishes logarithmically. But this
 may not be the case in the IR where we have very few criteria to check.
 In the model we are using we impose that the space is asymptotically AdS in the
 Einstein frame as this is the only choice consistent with the whole framework.}.

The coupling constant is also fixed to leading order from the coupling
 of the dilaton to $D_3$ branes (up to an overall multiplicative factor).
 Subleading (in perturbation theory) redefinitions of the coupling constant
 and the energy lead to changes in the $\beta$-function beyond two loops.

More in detail, as it  has been described in \cite{kirihqcd,kirdiss},
the general form of the kinetic term for the gauge fields on a $D_3$ brane is expected to be:
\be
S_{F^2}=e^{-\Phi}Z(R,\xi)Tr[F^2]\sp \xi\equiv -e^{2\Phi}{F^2_5\over 5!}
\ee
where $Z(R,\xi)$ is an (unknown) function of curvature $R$ and the five-form field strength, $\xi$.
At weak background fields, $Z\simeq -{1\over 4}+\cdots$.
In the UV regime, expanding near the boundary in powers of the coupling $\lambda \equiv N_c e^{\Phi}$ we obtain, \cite{kirdiss}
\be
S_{F^2}=N_c~Tr[F^2]~{1\over \lambda}\left[Z(R_*,\xi_*)-{Z_{\xi}(R_*,\xi_*)\over F_{\xi\xi}(R_*,\xi_*)\sqrt{\xi_*}}
{\l\over \ell}+{\cal O}(\l^2)\right]
\ee
where $F(R,\xi)$ is the bulk effective action and $R_*,\xi_*$ are the boundary values for these parameters.
Therefore the true 't Hooft coupling of QCD is
\be
\lambda_{\rm 't ~Hooft}=-{\l\over Z(R_*,\xi_*)}\left[1+{Z_{\xi}(R_*,\xi_*)\over Z(R_*,\xi_*)F_{\xi\xi}(R_*,\xi_*)\sqrt{\xi_*}}
{\l\over \ell}+{\cal O}(\l^2)\right] \;.
\ee
In the IR, more important changes can appear between our $\l$ and other definitions as for example in lattice calculations.

In the region of strong coupling we know much less
 in order to be guided concerning the correct definition of the energy.
We can obtain some hints however by comparing with lattice
results.\footnote{We would like to thank K. Kajantie for asking the question, suggesting to compare with lattice
 data, and providing the appropriate references.} In particular, based on lattice calculations using
the Schr\"ondiger functional \index{Schr\"ondiger functional} approach \cite{kirsommer}, it is argued that at long
distance $L$ the 't Hooft coupling constant scales as \be \l_{\rm
lat}\sim e^{m L}\sp m\simeq {3\over 4}m_{0^{++}} \;. \ee This was
based on a specific definition of the coupling constant, and
length scale on the lattice as well as on numerical data, and some
general expectations on the fall-off of correlations in a massive
theory. This suggests an IR $\beta$ function of the form
\be
L{d\l\over dL}={\l}\log{\l\over \l_0}\sp \l=\l_0 ~e^{mL} \;.
\ee

On the other hand our $\beta$-function at strong coupling uses the UV definition of energy,  $\log E=A_E$
 (the scale factor in the Einstein frame),  $E\sim 1/L$ and is
\be
L{d\l\over dL}={3\over 2}\l\left[1+{3\over 4}{a-1\over
a}{1\over \log\l}+\cdots\right]\sp \l\simeq ~\left({L\over L_0}\right)^{3\over 2} \;.
\label{kirbeta-n}
\ee
where $a$ is a parameter in the IR asymptotics of the potential.
The case we consider as best fitting YM is $a=2$ as then the
asymptotic glueball trajectories are linear..

Consider now taking as length scale the string
scale factor $e^{A_s}$ in the IR. \footnote{The string scale
factor is not a monotonic function on the whole manifold,
\cite{kirihqcd} and this is the reason that it was not taken as a
global energy scale. In particular in the UV, $e^{A_s}$ decreases
until it reaches a minimum. The existence of the minimum is
crucial for confinement. After this minimum $e^{A_s}$ increases
and diverges at the IR singularity.}  Since it increases, it
is consistent to consider it as a monotonic function of length.
{}From its relation to the Einstein scale factor $A_s=A_E+{2\over
3}\log \l$ and (\ref{kirbeta-n}) we obtain \be {d\l\over
dA_s}={2a\over a-1}\l\log\l +\cdots \;.\ee Therefore if we define as
length scale in the IR \be \log L={2a\over
a-1}A_s~~~\to~~~L=\left(e^{A_s}\right)^{2a\over a-1} \ee we obtain
a  running of the coupling compatible with the given lattice
scheme. Note however that $L=\left(e^{A_s}\right)^{2a\over a-1}$
cannot be a global choice but should be only valid in the IR. The
reason is that this function is not globally monotonic.

We conclude this section by restating that physical observables are independent of scheme.
But observables like the 't Hooft coupling constant do depend on schemes, and it is obvious
that our scheme is very different from lattice schemes in the IR.

\section{The potential and the parameters of the model}\label{kirparameters}

We will make  the following ansatz for  the potential\footnote{Further
studies of IHQCD\index{Improved Holographic QCD}  with different potentials can be found in \cite{kirkajantie2}.},
\be\label{kirpotential}
V(\l)  = {12\over \ell^2} \left\{ 1 + V_0 \l + V_1 \l^{4/3} \left[\log \left(1 + V_2 \l^{4/3} + V_3 \l^2\right) \right]^{1/2} \right\} ,
\ee
which interpolates between  the two asymptotic
behaviors (\ref{kirUVexp}) for small $\l$ and (\ref{kirIRexp}) for large $\l$, with $Q=2/3$ and $P=1/2$. Not
all the parameters entering this potential have physical relevance.
Below we will discuss the independent parameters of the model, and their physical meaning.

\paragraph{\bf The normalization of the coupling constant $\l$.} As discussed  in the previous section,
the relation between the bulk field $\l(r)$ and the physical QCD
't Hooft coupling $\l_t = g_{YM}^2 N_c$ is a priori unknown.
In the UV, the identification of the $D3$-brane coupling to the dilaton implies that
the relation is linear,  and depends on an {\em a priori} unknown coefficient $\kappa$,
defined as:
\be\label{kirlinear}
\l=\kappa\l_t.
\ee
The coefficient $\kappa$ can in principle be
identified by relating the perturbative UV  expansion of the Yang-Mills
$\beta$-function, to the holographic $\beta$-function \index{$\beta$-function!holographic}for the bulk field $\l$:
\bea
&&\beta(\l_t) = -\beta_0 \l_t^2 - \beta_1 \l_t^3 + \ldots \qquad \beta_0 = {22\over 3
(4\pi)^2}, \;\; \beta_1 = {51\over121}\, \beta_0^2 , \; \ldots  \label{kirbeta1}\\
&& \beta(\l) =  -b_0 \l^2 -b_1 \l^3 +\ldots, \qquad b_0 = {9\over 8} v_0, \; \; b_1
= \frac94 v_1 - \frac{207}{256}v_0^2 \,\,\ldots \;.\label{kirbeta2}
\eea

The two expressions (\ref{kirbeta1}) and (\ref{kirbeta2}) are consistent with a linear relation as in (\ref{kirlinear}),
 and expanding the identity  $\kappa \beta_t(\l_t) = \beta(\kappa \l_t)$  to lowest
order leads to:
\be\label{kirkappa}
\kappa = \beta_0/b_0.
\ee
Therefore, to relate the bulk field $\l$ to the true coupling $\l_t$ one looks at the linear term in
the expansion of the potential. More generally, the other $\beta$ function coefficients are related
by  $\b_n = \kappa^{n+1} b_n$, and the combinations   $b_n/b_0^{n+1}=\beta_n/\beta_0^{n+1}$ are $\kappa$-independent (however they are scheme-dependent for $n\geq 2$).

As discussed in Section \ref{kirscheme},
 the introduction of the coefficient $\kappa$ amounts  to a field redefinition and
therefore its precise value does not affect physical (scheme-independent)
 quantities. In this sense, $\kappa$ is not a parameter that can be fixed by matching some observable computed in the theory.
 Assuming the validity of
the relation (\ref{kirlinear}),  we could  eventually  fix $\kappa$
by matching  a RG-invariant (but scheme-dependent) quantity, e.g.
$\l$ at a given energy scale.

However, as we discuss later in this section, rescaling $\l$ in
the potential (thus changing $\kappa$)
 affects other parameters in the models,
 that are defined in the string frame, e.g. the fundamental string length $\ell_s$: if we hold
the physical QCD string tension fixed, the ratio $(\ell_s/\ell)$
scales with degree $-2/3$ under a rescaling of $\kappa$.

An important point to keep in mind, is that
 the simple linear relation (\ref{kirlinear}) may be modified at strong coupling, but
again this does not have any effect on physical observables.  {\em As long
as we compute RG-invariant and scheme-independent quantities, knowledge of
the exact relationship $\l = F(\l_t)$ is unnecessary. }

\paragraph{\bf The AdS scale \index{AdS scale} $\ell$.} This is set by
the overall normalization of the potential, and its choice  is
equivalent to fixing the unit of energy. It does not enter
dimensionless physical quantities.  As usual the AdS length at
large $N_c$ is much larger than the Planck length\index{Planck length} ($\ell_p \sim
1/(M_pN_c^{2/3})$, independently of  the 't Hooft coupling.

\paragraph{\bf The UV expansion coefficients of $V(\l)$.} They can be fixed order by order by matching
the Yang-Mills $\beta$-function.\index{$\beta$-function}  We impose  this matching up to
two-loops in the perturbative expansion, i.e. $O(\l^3)$ in
$\b(\l)$.  One could go to higher orders  by adding additional
powers of $\l$ inside the logarithm, but since our purpose is not
to give an accurate description of the theory in the UV, we choose
not to introduce extra parameters\footnote{Moreover, higher order
$\beta$-function coefficients are known to be scheme-dependent.}.

Identifying the energy scale with the Einstein frame scale factor, $\log E \equiv \log b(r)$,  we have
the relation (\ref{kirbeta2}) between the $\beta$-function coefficients  and the expansion parameters of $V(\l)$, with
\be
v_0 = V_0, \quad  v_1  = V_1 \sqrt{V_2}.
\ee
The term proportional to $V_2$ in eq. (\ref{kirpotential}) is needed to reproduce  the correct value
of the quantity $b_1/b_0^2 = \b_1/\b_0^2 =51/121$, which is invariant under rescaling of $\l$. Thus, $V_2$ is
not a free parameter, but  is fixed  in terms of $V_0$ and $V_1$ by:
\be\label{kirV2}
V_2 = b_0^4 \left({23  + 36\, b_1/b_0^2 \over 81 V_1 }\right)^2, \qquad b_0 = {9\over 8} V_0, \quad {b_1 \over b_0^2} = {51\over 121}.
\ee

As explained earlier in this section, when discussing the
normalization of the coupling, fixing the  coefficient $V_0$  is
the same as fixing  the normalization $\kappa$ through eq.
(\ref{kirkappa}). As we argued,  the actual value of $\kappa$ should
not have any physical consequences, so it is tempting to set
$V_0=1$ by a field redefinition, $\l \to \l/V_0$ and eliminate
this parameter altogether.

In fact, most
of the quantities we will compute are not sensitive to the value of $V_0$, but for certain quantities, such
as the string tension,  some extra care is needed.
In general, we can ask whether two models of the same
form (\ref{kira1}),  but  with different  potentials $V(\l)$ and  $\tilde{V}(\l)$, such
that $\tilde{V}(\l) = V(\alpha \l)$ for some constant $\alpha$,  lead to different physical predictions. As
we can change from one model to the other simply by a
 field redefinition $\l \to \alpha \l$ ( this
 has no effect on the other terms in the action in the Einstein frame,
eq. (\ref{kira1}) ), clearly the two potentials lead to the same result  for
any physical quantity that can be
 computed unambiguously from the Einstein frame action,
e.g. dimensionless ratios between glueball masses, critical temperature, latent heat etc.

However a rescaling of $\l$ does affect  the string frame metric,
since the latter  explicitly contains factors of $\l$: $b_s(r) = b(r) \l^{2/3}$  \cite{kirihqcd} thus,
under the rescaling $\l \to \alpha \l$,   $b_s(r) \to \alpha^{2/3} b_s(r)$.
 This means that any dimensionless ratio of two quantities, such that one of them remains
fixed in the  string frame and  the other in the Einstein frame,
will depend on $\alpha$. An example of this is the ratio $\ell_s/\ell$, where
$\ell_s$ is the string length, that we will discuss shortly.

Therefore, we  can safely perform a field redefinition and set $V_0$ to a given value,
as long as we are careful
when computing quantities that depend explicitly  on the fundamental string length.

Bearing this caveat in mind, we will choose a normalization such that $b_0 = \beta_0$, i.e.
\be\label{kirV0}
V_0= {8\over 9} \beta_0,
\ee
so that the normalization of $\l$ in the UV matches the physical Yang-Mills coupling. With this choice,
 out of the four free parameters $V_i$ appearing in (\ref{kirpotential})
only $V_1$ and $V_3$ play a non-trivial role ($V_2$ being fixed by eq. (\ref{kirV2})).

\paragraph{\bf The 5D Planck scale $M_p$.}\index{Planck scale}  $M_p$ appears in the overall normalization of the 5D action (\ref{kira1}). Therefore it
enters  the overall scale of quantities derived by evaluating the
on-shell action, e.g. the free energy and the black-hole mass. It
also sets the conversion factor between the entropy and horizon
area. $M_p$ cannot be fixed directly as we lack a detailed underlining string theory for YM.
 To obtain quantitative predictions, $M_p$  must be fixed in
terms of the other dimension-full quantity of the model, namely the
AdS scale $\ell$. As shown in \cite{kirGKMN2} this can be done by
imposing that the high-temperature limit of the black-hole free\index{black hole}
energy be that of a {\em free} gluon gas with the correct number of degrees of freedom\footnote{Note that this is conceptually different from the
${\cal N}=4$
case. There, near the boundary, the theory is strongly coupled and this number must be calculated in string theory.
 It is different by a factor of 3/4 from the free sYM answer. Here near the boundary the theory is free.
Therefore the number of degrees of freedom can be directly inferred.}. This
requires:
\be\label{kirPlanck} (M_p \ell)^3 = {1\over 45\pi^2}. \ee

\paragraph{\bf The string length.}\index{string length} In the non-critical approach the relation between the string length $\ell_s$ and the
5D Planck length (or the AdS length $\ell$)  is not known from first principles. The string length does not appear explicitly in the
2-derivative action (\ref{kira1}), but it enters quantities like the static quark-antiquark potential.\index{quark-antiquark potential}
The ratio  $\ell_s/\ell$  can be fixed phenomenologically to match the lattice results for the confining
string tension.

More in detail, the relation between the fundamental and the confining string tensions $T_f$ and $\sigma$ is given by:
\index{string tension}
\be\label{kirtension}
\sigma  = T_f  \, b^2(r_*) \l^{4/3}(r_*),
\ee
where $r_*$ is the point where the string frame scale factor, $b_s(r) \equiv b(r) \l^{2/3}(r)$, has
its minimum. Fixing the confining string tension by comparison with the lattice result we can
find $T_f$ (more precisely, the dimensionless quantity $T_f \ell^2$, since
the overall scale of the metric depends on $\ell$).
The string length is in turn  given
by  $\ell_s/\ell = 1/\sqrt{2\pi T_f \ell^2}$.

As is clear from eq. (\ref{kirtension}),  rescaling $\l \to \alpha \l$,  keeping  the value of the QCD string tension $\sigma$  and  of the AdS scale $\ell$ fixed, affects the fundamental string length  in AdS units
as $\ell_s/\ell \to  \alpha^{-2/3}(\ell_s/\ell) $. Therefore two models $a$ and $b$,  {\em defined in the Einstein frame} by eq. (\ref{kira1}), but with potentials related by $V_b(\l) = V_a(\alpha \l)$,
must have different fundamental string tensions in order to reproduce the same result for
the QCD string tension. The quantity $\ell_s/\ell$ therefore depends on the value of $V_0$.\\

\paragraph{\bf Integration constants.} Besides the parameters appearing directly
 in the gravitational action, there are also other physically relevant
 quantities that label different solutions to the 5-th order system of field equations.
Any solution is characterized by  a scale $\Lambda$, the
temperature $T$ and a value for the gluon condensate ${\cal G}$,
that correspond to three of the five independent integration
constants.\footnote{The remaining two are the value $f(0)$ which
should be set to one for the solution (\ref{kira7})  to obey the right UV
asymptotics, and an unphysical degree of freedom in the
reparametrization of the radial coordinate.}

Regularity at the horizon
fixes  ${\cal G}$ as a function of $T$, so that effectively the gluon condensate
is a temperature-dependent quantity.

The quantity $\Lambda$ controls the asymptotic form of the
solution, as it enters the dilaton running in the UV: $\l \simeq
-(b_0  \log r \Lambda )^{-1}$. It can be defined in a
reparametrization invariant way as: \be \label{kirLambda} \Lambda =
\ell^{-1} \lim_{\l\to 0} \left\{b(\l){\exp\left[ - {1\over
b_0\l}\right] \over \l^{b_1/b_0^2}} \right\}, \ee and it is fixed
once we specify the value of the scale factor $b(\l)$ at a given
$\l_0$.

Every choice of $\Lambda$
corresponds to an inequivalent class of solutions, that differ by UV  boundary conditions.
Each class is thermodynamically isolated, since solutions with different $\Lambda$'s have
infinite action difference. Thus, in the canonical partition sum we need to consider
only solutions with  a fixed value of $\Lambda$. However, this choice is merely a choice
of scale, as solutions with different $\Lambda$'s will give the same predictions
for any dimensionless quantity.
In short,  $\Lambda$ is the holographic dual  to the  QCD strong coupling scale: it
is defined by the initial condition to the holographic RG equations, and does not
affect dimensionless quantities such as mass ratios, etc. Therefore, as long
as all solutions we consider obey the same UV asymptotics, the actual value of $\Lambda$ is
immaterial, since the physical units of the system can always be set by fixing $\ell$.\\

To summarize, the only {\em nontrivial} {\it phenomenological} parameters we have at our disposal are $V_1$ and $V_3$ appearing
in (\ref{kirpotential}). The other quantities that enter our model are either fixed by the arguments
presented in this section, or they only affect trivially (e.g. by overall rescaling that can be absorbed in the definition of the
fundamental string scale)
the physical quantities.

In the next section we  present a numerical analysis of the solutions and thermodynamics
of the model defined by eq. (\ref{kirpotential}), and show that for an appropriate choice of the
parameters it reproduces the lattice results for the Yang-Mills deconfinement transition and
high-temperature phase as well as the zero temperature glueball data.

\section{Matching  the thermodynamics of  large-$N_c$ YM} \label{kirmatching}

Assuming a potential of the form (\ref{kirpotential}), we look for
values of the parameters such that the thermodynamics of the 5D
model match the lattice results for the thermodynamics of 4D YM.
As explained in Section \ref{kirscheme}, we set $V_0$ and $V_2$ as in eqs.
(\ref{kirV0}) and (\ref{kirV2}),
 respectively, with $b_0=\beta_0 = 22/3 (4\pi)^{-2}$.

We then vary  $V_1$ and $V_3$ only. We fix these parameters by
looking at thermodynamic quantities corresponding to the latent
heat per unit volume, and the pressure at one value of the temperature above
the transition, which we take as $2T_c$.

It is worth remarking
 that $V_1$ and $V_3$ are phenomenological parameters that we use to fit {\em dimensionless}
QCD quantities. The single (dimension-full) parameter of pure
Yang-Mills, the strong coupling scale, is an extra input that
fixes the overall energy  scale of our solution.

 Using the numerical method explained in \cite{kirfit}, for each set of
parameters $(V_1,V_3)$ we numerically generate black-hole solutions for  a range of values of $\l_h$,
then from the metric at the horizon and its derivative
 we extract the temperature and entropy functions $T(\l_h)$ and $S(\l_h)$, and the function ${\cal F}(\l_h)$
 from the integrated form of the first law,
 \be\label{kirintF}
{\cal F}(\l_h) = \int_{\l_h}^{+\infty}\intd\bar\l_h\,  S(\bar \l_h) {\intd T(\bar \l_h) \over \intd\bar \l_h} \quad.
\ee
 Here $S(\l_h)$ is given by (\ref{kirTS}) and both the large black hole\index{black hole!large}\index{black hole!small}
and small black-hole branches are needed in order the get the full
result for the free energy. This is because the integral in
(\ref{kirintF}) extends to $+\infty$, entering deeply in the small
black-hole branch.

\begin{figure}[h]
\begin{center}
\includegraphics[scale=0.8]{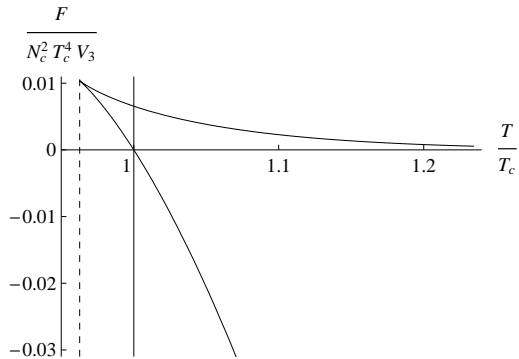}
\end{center}
 \caption[]{The Free energy density (in units of $T_{c}$) as a function of
 $T/T_c$, for $V_1=14$ and $V_3=170$. The vertical lines correspond to the
critical temperature (solid) and the minimum black-hole temperature (dashed). }
\label{kirFT}
 \end{figure}

\begin{figure}[h!]
 \begin{center}
\includegraphics[height=5cm,width=8cm]{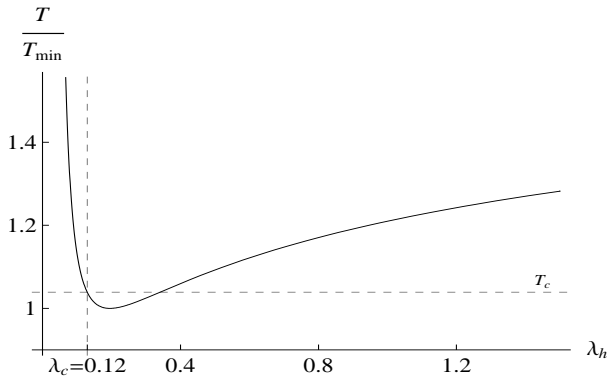}
\end{center}
 \caption[]{Temperature in units of $T_{min}$,  as a function of $\l_h$, for $V_1=14$ and $V_3=170$. The dashed
horizontal and vertical lines indicate the critical temperature and the critical value of the dilaton field at
at the horizon}\label{kirTfig}
\end{figure}

\begin{figure}[h!]
 \begin{center}
\includegraphics[height=5cm,width=8cm]{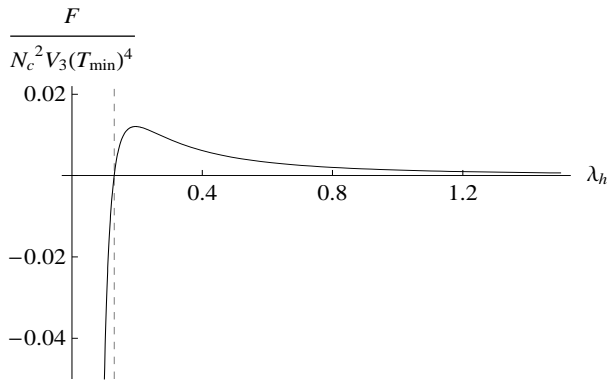}
\end{center}
 \caption[]{The free energy density in units of $T_{min}$,  as a function of $\l_h$}\label{kirFfig}
\end{figure}

\begin{figure}[h!]
 \begin{center}
\includegraphics[height=5cm,width=8cm]{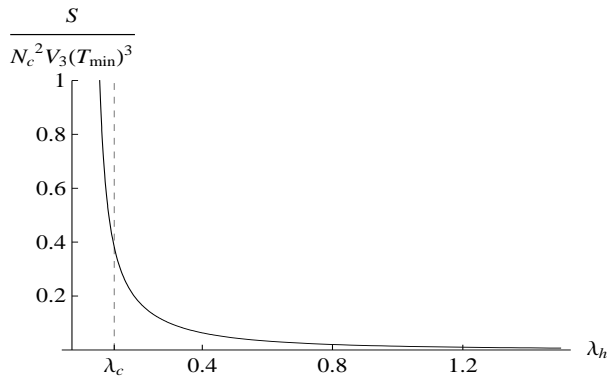}
\end{center}
 \caption[]{Entropy density  in units of $T_{min}$,  as a function of $\l_h$}
\label{kirSfig}
\end{figure}

The behavior of the thermodynamic  functions is  shown in Figures \ref{kirTfig}, \ref{kirFfig} and \ref{kirSfig}, for the best fit
parameter values that we discuss below.
 One  can see the existence of a minimal temperature $T_{min} = T(\l_{min})$,
 and a critical value $\l_c$ where ${\cal F}$ changes sign.
 The resulting function ${\cal F}(T)$
is shown in Figure \ref{kirFT}.

The phase transition\index{phase transition} is first order, and the latent heat per unit
volume $L_h$, normalized by $N_c^2 T_c^4$, is given by
the derivative of the curve  in Fig. \ref{kirFT} at $T/T_c=1$.
Equivalently, $L_h$ is proportional to the jump in the entropy
density $s=S/V_3$ at the phase transition from the thermal gas
(whose entropy is of $O(1)$, in the limit $N_c\to \infty$) to the
black hole (whose entropy scales as $N_c^2$ in the same limit):
thus, in the large $N_c$ limit, \be \label{kirlh} L_h \equiv T \Delta s
\simeq T_c s(\l_c) \ee
 up to terms of $O(1/N_c^2)$.

To fix  $V_1$ and $V_3$  we compare our results  to the data of G. Boyd et al. \cite{kirkarsch}.
 The relevant quantities to compare  are the dimensionless ratios $p(T)/T^4$, $e(T)/T^4$
and $s(T)/T^3$, where $p={\cal F}/V_3$ is the pressure, and $e = p
+ Ts$ is the energy density. Lattice results for these functions
are available in the range $T = T_c \sim 5T_c$, and can be seen in
Figure 7 of \cite{kirkarsch}. The analysis of \cite{kirkarsch}
correspond to  $N_c=3$, but one expects that the thermodynamic
functions do not change to much for large $N_c$\footnote{See e.g.
\cite{kirlucini}, in which results for $N_c=8$ do not different
significantly from those for $N_c=3$ as well as the recent high-precision data by Panero, \cite{kirpanero}.}.

An additional quantity of relevance is  the value for the ``dimensionless'' latent heat\index{latent heat} per unit volume, $L_h/T_c^4$
which for large $N_c$ was found in \cite{kirteperlucini} to be  $(L_h/T_c^4)_{lat} = 0.31 N_c^2$. The result
 for  $N_c^2=3$ is slightly lower ( $\simeq 0.28 N_c^2$).

As already noted in \cite{kirGKMN1,kirGKMN2}, the qualitative features of the thermodynamic functions
are generically reproduced in our setup: the curves $3 p(T)/T^4$, $e(T)/T^4$ and $3s(T)/4T^3$ increase
starting at $T_c$, then (very  slowly) approach the  constant free field value $\pi^2 N_c^2/15$ (given
by the Stefan-Boltzmann law) as $T$ increases.
By  computing the thermodynamic functions for various sets of  values of $V_1$ and $V_3$ we obtain that:

1) $V_1$ roughly controls the height reached by the curves $p(T)/T^4$, $e(T)/T^4$ and $s(T)/T^3$ at large $T/T_c$
($\sim$ a few): for larger $V_1$ the curves approach the free field limit faster;

2) $V_3$ does not
affect much the height of the curves at large $T/T_c$, but on the other hand it changes the latent heat, which is increasing as $V_3$
decreases.


The best fit corresponds to the values

\be\label{kirbestfit}
V_1= 14 \qquad
V_3 = 170.
\ee
Below we discuss the values of  various physical  quantities ( both related to thermodynamics, and to
zero-temperature properties) obtained with this choice of parameters.

\subsection{Latent heat and equation of state}

\begin{figure}[h]
 \begin{center}
\includegraphics[scale=1.0]{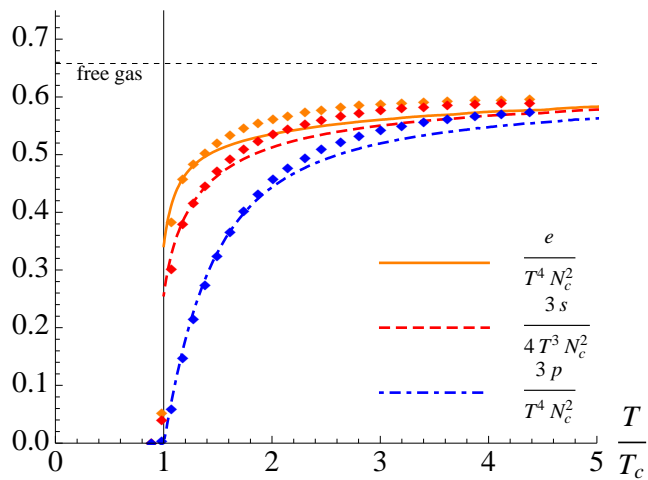}
\end{center}
 \caption[]{Temperature  dependence of the dimensionless thermodynamic densities
 $s/T^3$ (light blue), $p/T^4$ (dark blue) and $e/T^4$ (green),
  normalized such that they reach the common limiting value $\pi^2 /15$ (dashed
horizontal line) as $T\to \infty$. The dots correspond to the lattice data for $N_c=3$ \cite{kirkarsch}. }
\label{kiresp}
\end{figure}

The comparison between the curves $p(T)/T^4$, $e(T)/T^4$ and
$s(T)/T^3$
 obtained in our models with (\ref{kirbestfit}),   and the lattice results  \cite{kirkarsch}
  is shown in Figure \ref{kiresp}.
The match is remarkably good
for $T_c<T<2T_c$, and deviates slightly  from the lattice data in the range up to $5T_c$.

The latent heat we obtain is:
\be
 L_h/T_c^4 = 0.31 N_c^2,
\ee
which matches the lattice result for $N_c \to \infty$ \cite{kirteperlucini}.

\begin{figure}
 \begin{center}
\includegraphics[scale=0.8]{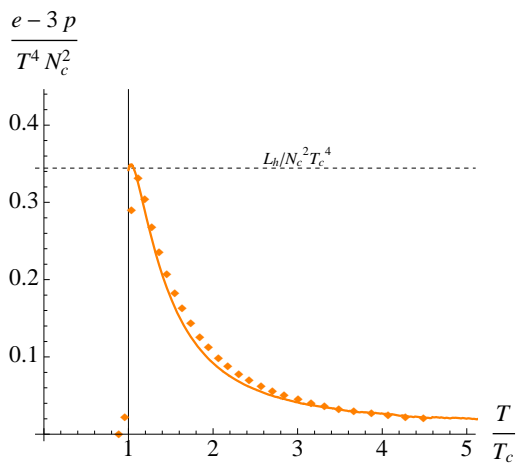}
\end{center}
 \caption[]{The trace anomaly as a function of temperature  in the deconfined phase
of the holographic model  (solid line) and the corresponding
lattice data \cite{kirkarsch} for $N_c=3$ (dots). The peak in the lattice data slightly above $T_c$ is expected to be
an artifact of the finite lattice volume. In the infinite volume limit the maximum value
of the curve  is at $T_c$, and it equals $L_h/N_c^2 T_c^4$.}
\label{kirtrace}
\end{figure}

\begin{figure}
 \begin{center}
\includegraphics[scale=0.5]{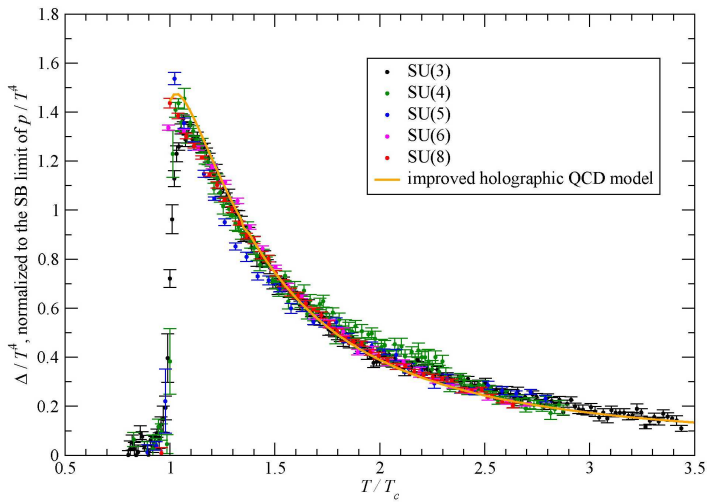}
\end{center}
 \caption[]{The rescaled trace anomaly (so that it is $N_c$-independent)
 as a function of temperature  in the deconfined phase
of the holographic model  (solid line) and the corresponding
recent high precision lattice data taken from \cite{kirpanero} for different $N_c$.
The errors shown are statistical only.
}
\label{kirtrace2}
\end{figure}

An interesting quantity is the {\em trace anomaly}\index{trace anomaly} $(e-3p)/T^4$,
(also known as {\em interaction measure})\index{interaction measure}, that indicates the
deviation from conformality, and it is proportional to the thermal gluon
condensate.\index{gluon condensate} The trace anomaly in our setup is shown, together with
the corresponding lattice data, in figure (\ref{kirtrace}), and the
agreement is again very good.
Our results agree even better with recent high-precision lattice calculations of the thermodynamics functions done by Panero at
different values of $N_c$ up to $N_c=8$, \cite{kirpanero}.
In figure \ref{kirtrace2} a comparison (taken from \cite{kirpanero})
of the normalized interaction measure with lattice results for different $N_c$ is shown.

\begin{figure}
 \begin{center}
\includegraphics[scale=0.8]{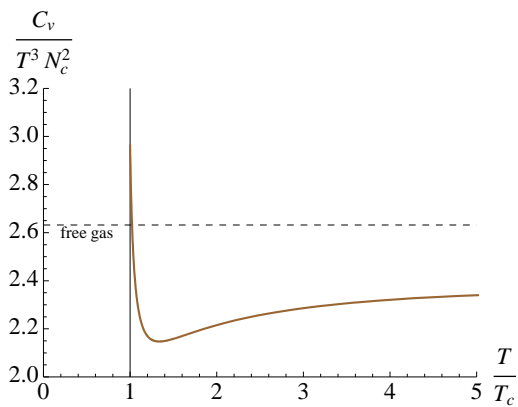}
\end{center}
 \caption[]{The specific heat (divided by $T^3$\textmd{}), as a function of temperature, in the
deconfined phase of the holographic model.}
\label{kircv}
\end{figure}

\begin{figure}
 \begin{center}
\includegraphics[scale=0.8]{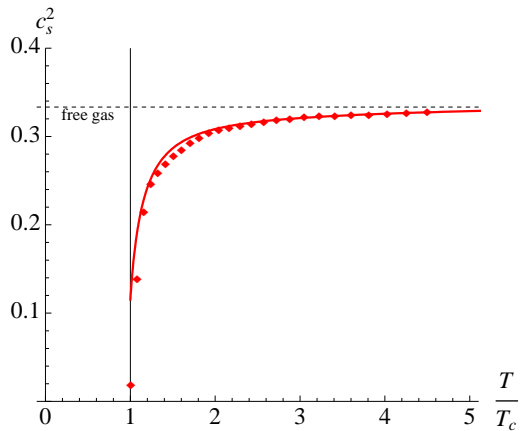}
\end{center}
 \caption[]{The speed of sound in the deconfined phase, as a function of temperature, for the
holographic model (solid line) and the corresponding   lattice data \cite{kirkarsch} for $N_c=3$ (dots).
The dashed horizontal line indicates the conformal limit $c_s^2 = 1/3$.}
\label{kirsound}
\end{figure}

We also compute the specific heat per unit volume $c_v$, and the
speed of sound $c_s$ in the deconfined phase,  by the  relations
\be\label{kircvcs} c_v = - T {\de^2 F \over \de T^2}, \qquad \quad
c_s^2 = {s\over c_v} \;. \ee These are shown in Figures \ref{kircv} and
\ref{kirsound} respectively. The speed of sound is shown together
with the lattice data, and the agreement is remarkable.

\subsection{Glueball spectrum}
\index{glueball spectrum}
In \cite{kirihqcd}, the single phenomenological parameters of the potential was fixed
by looking at the zero-temperature spectrum,
 i.e. by computing various glueball mass ratios and comparing
them  to the corresponding lattice results.
The masses are computed by deriving the effective action for the quadratic fluctuations around the background,
\cite{kirnit} and subsequently reducing the dynamics to four dimensions.

The associated thermodynamics for this potential was studied in \cite{kirGKMN1}
which was in qualitative agreement with lattice QCD results, but not in full quantitative agreement.
This is due to the fact that the thermodynamics depends more on the details
 of the potential than the glueball spectrum for the main Regge trajectories.
Here we use  the potential (\ref{kirpotential}), but with the two phenomenological parameters
$V_1$ and $V_3$ already determined by the thermodynamics
(\ref{kirbestfit}).

The glueball spectrum is obtained holographically as the spectrum
of normalizable fluctuations around the zero-temperature
background.
As explained in the introduction, and motivated in \cite{kirihqcd,kirdiss}, here we consider
explicitly  the 5D  metric, one scalar field
(the dilaton), and one pseudoscalar field (the axion).
 As a consequence, the only normalizable
fluctuations above the vacuum correspond to spin 0 and spin 2 glueballs\footnote{Spin 1 excitations of the
metric can be shown to be non-normalizable.} (more precisely, states
 with $J^{PC} = 0^{++}, 0^{-+}, 2^{++}$), each species containing an infinite discrete tower of excited
states.

In 4D YM there are many more operators generating glueballs, corresponding to different values of $J^{PC}$,
that are not considered here. These are expected to correspond holographically to other fields
in the noncritical string spectrum (e.g. form fields, which may yield  spin 1 and CP-odd spin 2 states)
and to higher string states that provide higher-spin glueballs. As the main focus is
in reproducing  the YM thermodynamics in detail rather than the entire glueball spectrum, we choose not to include
these states\footnote{A further  reason is that, unlike the scalar and (to some extent)
the pseudoscalar sector that we are considering, the action governing the higher Regge slopes
is less and less universal as one goes to higher masses. Only a precise knowledge of the underline string
theory is expected to provide detailed information for such states.}. Therefore we only compare the  mass spectrum
obtained in our model to the  lattice results for the lowest $0^{++}, 0^{-+}, 2^{++}$ glueballs and their
available excited states. These are limited to one for each spin 0 species, and none for the spin 2,
 in the study of \cite{kirchenetal}, which is the one we use for our  comparison.
This provides  two mass ratios in the CP-even sector and two  in the CP-odd sector.

The glueball masses are computed by first  solving numerically
the zero-temperature  Einstein's equations, by setting $f(r)=1$, and using the resulting metric
and dilaton to setup an  analogous  Schr\"odinger problem for the
fluctuations, \cite{kirihqcd}.
The results for the parity-conserving sector are shown in Table
\ref{kirmasses++}, and are in good agreement with those reported by
\cite{kirchenetal} for $N_c=3$, whereas the
results reported by \cite{kirteperlucini2} for large $N_c$ are somewhat larger.
 The CP-violating  sector (axial glueballs)\index{glueball spectrum!axial}  will be
discussed separately.

\begin{table}[h!]
\begin{center}
\caption{Glueball Masses}\label{kirmasses++}
\begin{tabular}{|r|c|c|c|}
\hline
& ~~~~HQCD~~~~~ & ~~~~~~$N_c=3$ \cite{kirchenetal}~~~~~~ & ~~~~~~~$N_c=\infty$ \cite{kirteperlucini2}~~~~~ \\
\hline\hline
$~~~m_{0^{*++}}/m_{0^{++}}~~~$ & 1.61 & 1.56(11) & 1.90(17)  \\
\hline
$m_{2^{++}}/m_{0^{++}}$ & 1.36 & 1.40(4) & 1.46(11)   \\
\hline \hline
\end{tabular}
\end{center}
\end{table}

We should add that there are other lattice studies (see e.g. \cite{kirmeyer}) that report
 additional excited states. Our mass ratios offer a somewhat worse
  fit of the mass ratios found in  \cite{kirmeyer} (whose results are
not entirely compatible with those of \cite{kirchenetal} for the states the two studies have in common).
We should stress however that  reproducing the detailed glueball spectrum
 is secondary here since  the main focus is  thermodynamics.
However, the comparison of our spectrum to the existing lattice results
shows that our model provides a good global fit to 4D YM also with respect to
quantities beyond thermodynamics.

Unlike the various  mass ratios,
the value of  any given mass in AdS-length units (e.g. $m_{0++} \ell$)  {\em does  depend}
 on the choice of integration constants in the UV, i.e.
on the value of $b_{UV}$ and $\l_{UV}$. Therefore its numerical
value does not have an intrinsic meaning. However it can be used
as a benchmark against which all other dimension-full quantities can
be measured (provided one always uses the same UV b.c. ).  On the
other hand, given a fixed set of initial conditions, asking that
$m_{0++}$ matches the physical value (in MeV) obtained on the
lattice, fixes the value of $\ell$ hence the energy unit.

\subsection{Critical temperature}

The thermodynamic quantities we have discussed so far, are
dimensionless ratios, in units of the critical temperature. To
compute $T_c$,   we need an extra dimension-full quantity which can
be used independently to set the unit of energy. In lattice
studies this is typically the confining string tension\index{string tension} $\sigma$ in
the $T=0$ vacuum, with a value of around $(440 MeV)^2$, and
results are given in terms of the dimensionless ratio
$T_c/\sqrt{\sigma}$. In our case we cannot compute $\sigma$
directly, since it depends on the {\em fundamental} string
tension, which is a priori unknown. Instead, we take the mass
$m_{0++}$ of the lowest-lying glueball state as a reference.

We compute  $m_{0++}$ with  the potential  (\ref{kirpotential}), with
$V_1$ and $V_3$ fixed as in (\ref{kirbestfit}),  then compare
$T_c/m_{0++}$ to the same quantity obtained on the lattice. For
the  lattice result, we  take the large $N_c$ result of
\cite{kirteperlucini},  $T_c/\sqrt{\sigma} = 0.5970(38)$, and combine
it with the large $N_c$ result for the lowest-lying glueball  mass
\cite{kirteperlucini2},  $m_{0++}/\sqrt{\sigma}= 3.37(15)$. The two
results are in fair  agreement, without need to adjust any extra
parameter: \be \left(T_c\over m_0\right)_{hQCD}  = 0.167, \qquad
\left(T_c\over m_0\right)_{lattice}  = 0.177(7) \;.\ee In physical
units, the critical temperature we obtain is given by \be T_c =
0.56\, \sqrt{\sigma}  =  247 \, MeV. \ee

\subsection{String tension}

 The fundamental string tension $T_{f}={1\over 2\pi\ell_s^2}$ cannot
 be computed from first principles in our model, but can be
obtained using as extra input the lattice value of the confining
string tension $\sigma$, at $T=0$. The fundamental and
confining string tensions are related by eq. (\ref{kirtension}).
\index{string tension}

As
for the critical temperature, we can relate $T_f$ to the value of the lowest-lying glueball mass,
by using the lattice relation  $\sqrt{\sigma} = {m_{0++}\over 3.37 } $ \cite{kirteperlucini2}. Since what
we actually compute numerically is $m_{0++} \ell$, this allows us to obtain the string tension $T_f$ (and fundamental
string length $\ell_s = 1/\sqrt{2\pi T_f}$ in AdS units:
\be\label{kirtension-numbers}
T_f  \ell^2 = 0.19,  \qquad \ell_s/\ell = 0.15 \;.
\ee
This shows that the fundamental string length\index{string length}
 in our model is about an order of magnitude smaller than the  AdS length.
The meaning of this fact is a little  more complicated conceptually, as
the discussion in \cite{kirdiss} indicates. Also, we should
stress that, as discussed in Section \ref{kirparameters}, this
result depends on our choice of the overall normalization of $\l$:
changing the potential
 by $\l \to \kappa \l$ will yield different
 numerical values in (\ref{kirtension-numbers}) without affecting
the other physical quantities.

Another related observable is the spatial string tension.\index{string tension!spatial}
It is calculated from the expectation value of the rectangular Wilson loop
which  stretches in spatial dimensions only. This has been calculated on
 the lattice \cite{kirspatial}, as well as using the
high-temperature (resumed) perturbative expansion plus a
zero-temperature calculation of the string tension in three-dimensional YM theory, \cite{kirlaine}. The two calculation agree reasonably well.

The spatial string tension at finite temperature can be calculated in IHQCD, \cite{kirkajantie1} by calculating the relevant Wilson loop.
Very good agreement was found with the lattice calculations, especially at temperatures not far from the phase transition.

Finally, several calculations of quark-antiquark potentials \index{quark-antiquark potential} exist. At zero temperature the long distance asymptotics of the quark potential
was calculated in \cite{kirihqcd} and used to classify the dilaton potentials as a function of the confinement property.
The full quark potential including the short distance behavior was computed in \cite{kirzeng}. There a comparison to the Cornell potential was done
as well as with quarkonium spectra finding excellent agreement with data.\index{quarkonium}
The issue of quarkonium potentials from IHQCD-like theories was also recently discussed in \cite{kirpirner}.

Finally the Polyakov loop was recently computed \cite{kirnoronha}
in similar Einstein dilaton models that were studied first in \cite{kirgubser}.

\subsection{CP-odd sector}

The CP-odd sector of pure Yang-Mills is described holographically
by the addition of a bulk  pseudoscalar field $a(r)$ (the {\em
axion})\index{axion} with action:\footnote{This action was justified in \cite{kirihqcd,kirdiss}.
 The dilaton dependent coefficient $Z(\l)$ is
encoding both the dilaton dependence as well as the UV curvature
dependence of the axion kinetic terms in the associated string theory.
We cannot determine it directly from the string theory,
but we pin it down by a combination of first principles and lattice input, as we explain further below.}
\be\label{kiraxionaction} S_{axion} =
{M_p^3\over2} \int d^5 x  Z(\l) \sqrt{-g} (\de^\mu a) (\de_\mu a) \;.
\ee
The field $a(r)$ is dual to the topological density operator
$\tr F\tilde{F}$. The prefactor $Z(\l)$ is  a dilaton-dependent
normalization. The axion action is suppressed by a factor
$1/N_c^2$ with respect to the action (\ref{kira1}) for the dilaton and the
metric, meaning that in the large-$N_c$ limit one can neglect
the back-reaction of the axion on the background.

As shown in \cite{kirihqcd}, requiring the correct scaling  of $a(r)$
in the UV, and phenomenologically  consistent axial glueball
masses, constrain the asymptotics of $Z(\l)$ as follows:
\be\label{kirZ1} Z(\l)  \sim Z_0 \;,\; \l \to 0; \qquad Z(\l) \sim \l^4
\;,\;\l \to \infty, \ee where $Z_0$ is a constant.
 As a simple interpolating function between
these large- and small- $\l$ asymptotics we can take the
following: \be\label{kirZ2} Z(\l)  = Z_0(1 + c_a \l^4). \ee The
parameter $Z_0$ can be fixed by matching the topological
susceptibility of pure Yang-Mills theory,
 whereas $c_a$ can be fixed by looking at the axial glueball mass spectrum.

\paragraph{\bf Axial glueballs.}
As in \cite{kirihqcd}, we can fix $c_a$ by matching to the
lattice results the mass ratio $m_{0-+}/m_{0++}$ between the lowest-lying axial and scalar glueball states.
This is independent of the overall coefficient $Z_0$ in (\ref{kirZ2}). The lattice value $m_{0-+}/m_{0++} = 1.49$ \cite{kirchenetal}
is obtained for:
\be\label{kirca}
c_a = 0.26.
\ee
  With this choice, the mass of the first excited axial glueball  state
is in good agreement  with the corresponding lattice result
\cite{kirchenetal}: \be \left({m_{0-+*}\over m_{0++}}\right)_{hQCD} =
2.10 \qquad \left({m_{0-+*}\over m_{0++}}\right)_{lattice} =
2.12(10) \;.\ee

\paragraph{\bf Topological Susceptibility.}\index{topological susceptibility}
In pure Yang-Mills, the topological $\chi$  susceptibility is
defined by: \be\label{kirchi1} E(\theta) = {1\over 2}\, \chi \,
\theta^2, \ee where $E(\theta)$ is the vacuum energy density in
presence of a $\theta$-parameter. $E(\theta)$ can be computed
holographically by solving for the  axion profile $a(r)$ on a
given background, and evaluating the action  (\ref{kiraxionaction})
on-shell.

In the deconfined phase, the axion profile is trivial, implying a vanishing
topological susceptibility \cite{kirGKMN2}. This is in agreement
with large-$N_c$ arguments and lattice results \cite{kirltheta}.

\begin{figure}[h]
 \begin{center}
\includegraphics[scale=0.8]{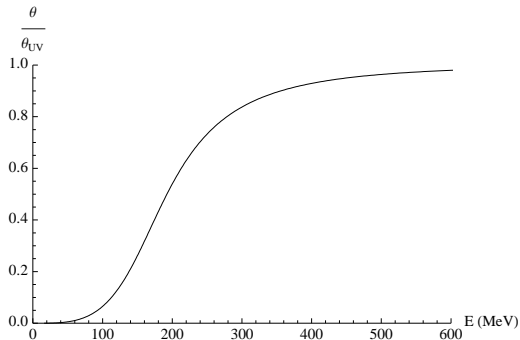}
\end{center}
 \caption[]{Axion profile in the radial direction. The  $x$-axis is
taken to be the energy scale, $E(r)=E_0 b(r)$, where the unit $E_0$ is
fixed to match the lowest glueball mass.}
\label{kiraxionprofile}
\end{figure}

In the low-temperature phase, the axion acquires a non-trivial profile,
\be\label{kiraprofile}
a(r) = a_{UV} \, {F(r)\over F(0)}, \qquad F(r)\equiv \int_r^{\infty} {dr \over Z(\l(r)) e^{3A(r)} } \;.
\ee
This profile is shown, for the case at hand, in Figure \ref{kiraxionprofile}, where
the axion is normalized to its UV value.

The topological susceptibility is given by \cite{kirihqcd}:
\be\label{kirchi2} \chi = M_p^3 F(0)^{-1} = M_p^3\left[ \int_0^\infty
{dr \over e^{3A(r)}Z(r)} \right]^{-1}, \ee where $Z(r)\equiv
Z(\l(r))$. Evaluating this expression numerically with $Z(\l)$ as
in (\ref{kirZ2}),  and $c_a=0.26$ (to match the axial glueball
spectrum ),  we can determine the coefficient $Z_0$ by looking at
the lattice result for $\chi$. For $N_c=3$,  \cite{kirDelDebbio}
obtained $\chi = (191 MeV)^4$, which requires $Z_0 = 133$.

In Table \ref{kirparameters} we present a summary of the various
physical quantities discussed in this section, as obtained in our
holographic  model, and their comparison with the lattice results
for large $N_c$ (when available) and for $N_c=3$. The quantities
shown in the upper half of the table  are the ones that were used
to fix the free parameters (reported in the last column)  of the
holographic model.

\subsection{Coupling normalization}

Finally, we can relate the field $\l(r)$ to the running 't Hooft
coupling. All other quantities  we have discussed so far are
scheme-independent and RG-invariant. This is not the case for the
identification of the physical YM 't Hooft coupling,  which is
scheme dependent.

\begin{figure}[h]
 \begin{center}
\includegraphics[scale=0.8]{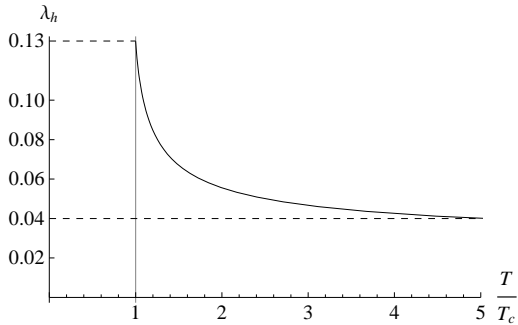}
\end{center}
 \caption[]{The coupling at the horizon as a function of  temperature in the range $T_c$--$5T_c$.}
\label{kircouplingvsT}
\end{figure}

In the black-hole phase we can take $\l_h \equiv \l(r_h)$ as a measure of the
temperature-dependent coupling. In figure \ref{kircouplingvsT} we show $\l_h$ as a function of the temperature
in the range $T_c$ to  $5T_c$.

As a reference, we may take the result of \cite{kirkarsch}, that
found $g^2(5T_c)\simeq 1.5$ for $N_c=3$, which translates to
$\l_t(5T_c) \simeq 5$. On the other hand,  if we  make the
assumption that the  identification $\l = \l_t$  is valid at all
scales (not only in the UV), we find in our model $\l_t(5T_c)
\simeq 0.04$ (see Figure \ref{kircouplingvsT}), i.e. a factor of 100
smaller than the lattice result.

 This discrepancy is almost certainly due to the identification
(\ref{kirlinear})being very different from lattice at strong coupling.

\begin{table}
\caption{Collected in this table is the complete set of physical quantities
 that we computed in our model and compared with data.
The upper half of the table contains the quantities that we used as input
 (shown in boldface) for the holographic QCD model (HQCD). Each quantity\index{Improved Holographic QCD}
 can be roughly associated to one parameter of the model (last column). The lower half of the table
contains our ``postdictions'' (i.e. quantities that we computed after all
the parameters were fixed) and the
comparison with the corresponding lattice results.  The value we find for
 the critical temperature corresponds to $T_c=247 MeV$.
}
\begin{center}
\begin{tabular}{|r|c|c|c|l|}
\hline
& HQCD &  lattice $N_c=3$  & lattice $N_c\to \infty$ & Parameter \\
\hline \hline
&&&&\\
$[p/(N_c^2 T^4)]_{T=2T_c}$ & {\bf 1.2} & {\bf 1.2} & -  & $V1=14$\\
&&&&\\
$L_h/(N_c^2 T_c^4)$ & {\bf 0.31} & 0.28 \cite{kirkarsch} & {\bf 0.31} \cite{kirteperlucini} & $V3=170$ \\
&&&&\\
$[p/(N_c^2 T^4)]_{T\to +\infty}$ & {\bf $\pi^2/45$} & {\bf $\pi^2/45$}  &{\bf $\pi^2/45$ }& $M_{p} \ell=[45 \pi^2]^{-1/3}$ \\
&&&&\\
$m_{0^{++}}/\sqrt{\sigma}$ & {\bf 3.37}  & 3.56 \cite{kirchenetal} &{\bf  3.37} \cite{kirteperlucini2}&$\ell_s/\ell =0.15$\\
&&&&\\
$m_{0^{-+}}/m_{0^{++}}$ & {\bf 1.49} &  {\bf 1.49} \cite{kirchenetal}  &- & $c_a= 0.26 $\\
&&&&\\
$\chi$ & {\bf $(191 MeV)^4$} &  $ (191 MeV)^4$ \cite{kirDelDebbio} & - & $Z_0=133$ \\
&&&&\\
\hline\hline
&&&&\\
$T_c/m_{0^{++}}$ & 0.167 &- &0.177(7) &  \\
&&&&\\
\hline
&&&&\\
$m_{0^{*++}}/m_{0^{++}}$ &1.61 &1.56(11) &1.90(17) & \\
&&&&\\
$m_{2^{++}}/m_{0^{++}}$ & 1.36 & 1.40(4)& 1.46(11)& \\
&&&&\\
\hline
&&&&\\
$m_{0^{*-+}}/m_{0^{++}}$ & 2.10 & 2.12(10) & - & \\
&&&&\\
\hline
\hline
\end{tabular}
\label{kirresults}
\end{center}
\end{table}

\section{Bulk viscosity}\label{kirBULKVISC}
\index{viscosity!bulk}

The bulk viscosity $\zeta$ is  an important probe of the quark-gluon plasma.
Its profile as a function of T reveals information
regarding the dynamics of the phase transition. In particular, both from the
low-energy theorems and lattice studies \cite{kirViscos1,kirViscos2,kirViscos3}, there is evidence
that $\zeta$ increases  near $T_c$.

For a viscous fluid the shear $\eta$ and bulk $\zeta$ viscosities are defined via the rate of entropy production as
\be
{\partial s\over\partial t}={\eta\over T}\left[\partial_iv_j+\partial_jv_i-{2\over 3}(\partial\cdot v)\delta_{ij}\right]^2
+{\zeta\over T}(\partial\cdot v)^2
\ee

Therefore, in a holographic setup, the bulk viscosity can be defined
as the response of the diagonal spatial components of the
stress-energy tensor to a small fluctuation of the metric.
It can be directly related to the retarded Green's
function of the stress-energy tensor by Kubo's linear response
theory:
\be \label{kirzeta1} \z=-\frac19 \lim_{\omega \rightarrow 0}
\frac{1}{\omega} Im G_R(\omega,0),
\ee
 where $G_R(w,\vec{p})$ is the
Fourier transform of  retarded Green's function of the
stress-energy tensor:
\be\lab{kirgreen} G_R(w,\vec{p}) = -i\int
d^{3}x dt e^{i \omega t - i \vec{p}\cdot\vec{x}}\theta(t)
\sum^{3}_{i,j=1}\langle\,[T_{ii}(t,\vec{x}),T_{jj}(0,0)]\,\rangle. \ee
A direct
computation of the RHS on the lattice is non-trivial as it
requires analytic continuation to Lorentzian space-time. In refs.
\cite{kirViscos1},\cite{kirViscos2}  the low energy theorems of QCD, as well as (equilibrium) lattice data at
finite temperature were used in order to evaluate a particular moment of the
spectral density of the relevant correlator.
using a parametrization of the spectral density via two time-dependent constants, one of which is the bulk viscosity
a relation for their product was obtained as a function of temperature.
This can be converted to a relation for $\zeta$, assuming the other constant varies slowly with temperature.

The conclusion was that  $\zeta/s$
increases near $T_c$. Another conclusion is that the fermionic contributions
 to $\zeta$ are small compared to the glue contributions.

The  weak point of the approach of
\cite{kirViscos2}, is that  it requires an ansatz on the spectrum of
energy fluctuations, and further assumptions on the other parameters.
which are not derived from first principles.

A direct lattice study of the bulk viscosity was also made in
\cite{kirViscos3}. Here, the result is also qualitatively similar
\ref{kirViscosFigI}. However, the systematic errors in this computation are large
especially near $T_c$,  mostly due to the analytic continuation that
one has to perform after computing the Euclidean correlator on the
lattice.

The results of references \cite{kirViscos1},\cite{kirViscos2} and the
assumptions of the lattice calculation have been recently challenged in
\cite{kirms}.

\subsection{The holographic computation}

The holographic approach offers a new way of computing the bulk viscosity.
 In the
holographic set-up, $\zeta$ is obtained from (\ref{kirzeta1}). Using
the standard AdS/CFT prescription, the two point-function of the
energy-momentum tensor can be read off from the asymptotic
behavior of the metric perturbations $\delta g_{\mu \nu}$. This is
similar in spirit to the holographic computation of the shear
viscosity \cite{kirShear}, but it is technically more involved. A
recent treatment of the fluctuation equation governing the scalar
mode  of a general  Einstein-Dilaton system can be found in \cite{kirspringer}.
Here, we  follow the method proposed by \cite{kirGubser:2008sz}.

As explained in \cite{kirGubser:2008sz}, one only needs to examine
the equations of motion in the gauge $r=\Phi$, where the radial
coordinate is equal to the dilaton. In our type of metrics, the applicability of this
method requires some clarifications, that we provide  in \cite{kirdrag}.
Using $SO(3)$ invariance
and the five remaining gauge degrees of freedom the metric
perturbations can be diagonalized as
\be\label{kirpert1} \delta g =
diag(g_{00},g_{11},g_{11},g_{11},g_{55}), \ee
where
\be\label{kirpert2} g_{00} = -e^{2A}f [1+ h_{00}(\f)e^{-i\o t}],\quad
g_{11} = e^{2A} [1+ h_{11}(\f)e^{-i\o t}],\quad ,\ee
$$g_{55} =
\frac{e^{2B}}{f}[1+ h_{55}(\f)e^{-i\o t}],
$$
 where the functions $A$ and $B$ emerge from the metric
 \be ds^2 =e^{2A(\f)}(-f dt^2 +dx_mdx^m) + e^{2B(\f)}\frac{d\f^2}{f}.
 \label{kirBH2}\ee
Here, the fluctuations are taken to be harmonic functions of $t$ while having an
arbitrary dependence on $\f$.

The bulk viscosity depends only on the correlator of the diagonal
components of the metric and so it suffices to look for the
asymptotics of $h_{11}$.  Interestingly, in the $r=\Phi$ gauge
this decouples from the other components of the metric and
satisfies the following equation\footnote{Difference in the
various numerical factors in this equation w.r.t
\cite{kirGubser:2008sz} is due to our different normalization of the
dilaton kinetic term.} \be \label{kirhw} h_{11}''
-\left(-\frac{8}{9A'}-4A'+3B'-\frac{f'}{f}\right)h_{11}'
-\left(-\frac{e^{2B-2A}}{f^2}\omega^2
+\frac{4f'}{9fA'}-\frac{f'B'}{f}\right)h_{11} = 0\;. \ee One needs
to impose two boundary conditions. First, we require that only the
infalling condition survives at the horizon: \be \label{kirinfall}
h_{11} \to c_b (\f_h-\f)^{-\frac{i\omega}{4\pi T}}, \qquad
\f\to\f_h, \ee where $c_b$ is a normalization factor. The second
boundary condition is that $h_{11}$ has unit normalization on the
boundary: \be\label{kirunitnorm} h_{11}\to 1, \qquad \f\to-\infty.
\ee Having solved for $h_{11}(\f)$, Kubo's formula (\ref{kirzeta1})
and a wise use of the AdS/CFT prescription to compute the
stress-energy correlation function \cite{kirGubser:2008sz} determines
the ratio of bulk viscosity as follows.

The AdS/CFT prescription relates the imaginary part of the retarded\index{retarded Green's function}
$T_{ii}$ Green's function to the number flux of the $h_{11}$
gravitons ${\cal F}$ \cite{kirGubser:2008sz}:
\be\lab{kirgreen2} Im\,
G_R(\omega,0) = -\frac{\cal F}{16\pi G_5} \ee where the flux can be
calculated as the Noether current\index{Noether current}  associated to the $U(1)$
symmetry $h_{11}\to e^{i\theta} h_{11}$ in the gravitational
action for fluctuations. One finds, \be\lab{kirflux1} {\cal F}  =
i\frac{e^{4A-B}f}{3A^{'2}}[h_{11}^*h_{11}'-h_{11}h_{11}^{*'}]. \ee
As ${\cal F}$ is independent of the radial variable, one can compute it at
any $\f$, most easily near the horizon, where $h_{11}$ takes the
form (\ref{kirinfall}). Using also the fact that $(dA/d\f )(\f_h) =
-8V(\f_h)/9V'(\f_h)$, one finds
\be
\lab{kirflux2} {\cal F(\o)} = \frac{27}{32} \o |c_b(\o)|^2
e^{3A(\f_h)}\frac{V'(\f_h)^2}{V(\f_h)}.
\ee
Then, (\ref{kirzeta1})
and (\ref{kirgreen2}) determine the ratio of bulk viscosity and the
entropy density as,
\be
\label{kirzs} \frac{\zeta}{s} =
\frac{3}{32\pi} \left(\frac{V'(\f_h)}{V(\f_h)}\right)^2 |c_b|^2.
\ee
In the derivation we use the Bekenstein-Hawking formula for\index{Bekenstein-Hawking formula}
the entropy density, $s = \exp{3A(\f_h)}/4G_5$.

To find $\zeta$ we need to find $c_b$ only in the limit $\omega\to 0$.
The computation is performed by numerically solving equation (\ref{kirhw})
with the appropriate boundary conditions.
There are two separate methods that one can employ to
determine the quantity $c_b$:
\begin{enumerate}
\item
One can solve (\ref{kirhw}) numerically with a fixed $\omega/T$, but small enough so
that $c_b$ reaches a fixed value. The method is valid also for
finite values of $\omega$. From a practical point of view, it is easier
to solve (\ref{kirhw}) with the boundary condition (\ref{kirinfall})
with a unit normalization factor, read off the value on the
boundary $h_{11}(-\infty)$ from the solution and finally use the
symmetry of (\ref{kirhw}) under constant scalings of $h_{11}$ to
determine $|c_b| = 1/|h_{11}(-\infty)|$.

\item An alternative method of computation that directly extracts
the information at $\omega=0$ follows from the following trick
\cite{kirGubser:2008sz}. Instead of solving (\ref{kirhw}) for small but
finite $\omega$, one can instead solve it for $\omega=0$. This is a simpler
equation, yet complicated enough to still evade analytic solution. Let
us call this solution $h_{11}^0$.
 One numerically solves it by fixing the boundary conditions on the boundary:
$h^0_{11}(-\infty)=1$ and the derivative $dh^0_{11}/d\f (-\infty)$ is
 chosen such that $h_{11}$ is regular at the horizon.
Matching this solution to the expansion of (\ref{kirinfall}) for small $\omega$
than yields $|c_b| = h^0_{11}(\f_h)$.
\end{enumerate}

\begin{figure}
 \begin{center}
\includegraphics[height=6cm,width=8cm]{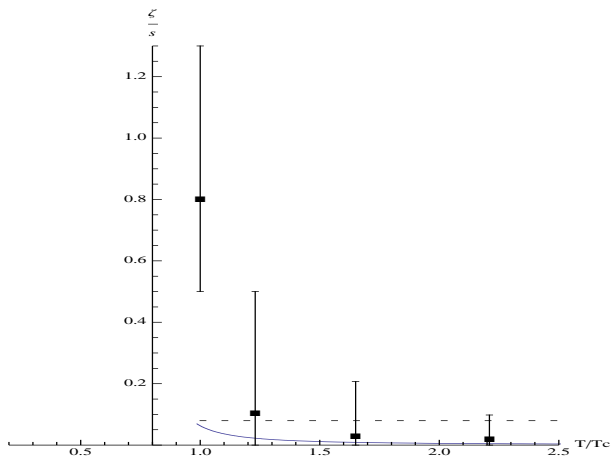}
 \end{center}
 \caption[]{Plot of $\zeta/s$ (continuous line) calculated in Improved Holographic QCD model.
  This is compared with the lattice data of \cite{kirViscos3}
  that are shown as boxes. The horizontal dashed line is
  indicating the (universal) value of ${\eta\over s}$ for comparison. }
\label{kirViscosFigI}
 \end{figure}

Both methods were used to obtain $\zeta/s$ as a function of T and
checked that they yield the same result. As explained in
\cite{kirGKMN2}, most of the thermodynamic observables are easily
computed using the method of scalar variables, \cite{kirGKMN2,kirdrag}.

The results are
presented in figure \ref{kirViscosFigI}. This figure gives a comparison
of the curve  obtained by the holographic calculation sketched above
by  solving (\ref{kirhw}) and the lattice data of \cite{kirViscos3}. We also
show $\eta/s=1/4\pi$ in this figure for comparison. The result is
qualitatively similar to the lattice result where $\zeta/s$
increases as $T$ approaches $T_c$, however the rate of increase is
slower than the lattice. As a result, we obtain a value
$\zeta/s(T_c)\approx 0.06$ that is an order of magnitude smaller
than the lattice result\cite{kirViscos3} which is 0.8. Note however
that the error bars in the lattice evaluation are large near $T_c$
and do not include all possible systematic errors from the
analytic continuation.

We should note the fact that the holographic calculation gives a
 smaller value for the bulk viscosity near $T_c$
than the lattice calculation is generic and has been found for
other potentials with similar IR asymptotics,
\cite{kirGubser:2008sz}. The fact that the value of $\z/s$ near $T_c$ is
 correlated with the IR asymptotics of the potential  will
be shown further below.

Another fact that one observes from figure \ref{kirViscosFigI} is
that $\zeta/s$ vanishes in the high $T$ limit. This reflects the
conformal invariance in the UV and can be shown analytically as
follows. $\z/s$ is determined by formula (\ref{kirzs}). In the high T
limit, (corresponding to $\l_h\to 0$, near the boundary), the
fluctuation coefficient $|c_b|\to 1$. This is because of the
boundary condition $h_{11}(\l=0)=1$. We use the relation between T
and $\l_h$ in the high T limit
\cite{kirGKMN2},
\be
\lab{kirlhThighT}
\l_h\to \le(\beta_0\log(\pi T/\Lambda)\ri)^{-1}.
\ee
Substitution
in (\ref{kirzs}) leads to the result,
\be \lab{kirhighTbBH}
\frac{\z}{s}\bigg|_{\rm large}\to \frac{2}{27\pi}\frac{1}{\log^2(\pi
T/\Lambda)}, \qquad as\,\,\, T\to\infty. \ee As $s$ itself
diverges as $T^3$ in this limit -- it corresponds to an ideal gas
--  we learn  that $\zeta$ also diverges as $T^3/\log^2(T)$.
Divergence at high T is expected from the bulk-viscosity of an
{\em ideal gas}. We do not expect however the details of the
asymptotic result to match  with the pQCD result, for the same
reasons that the shear-viscosity-to-entropy ratio does not,
\cite{kirdiss}. We note however, that the asymptotic T-dependence
is very similar to the pQCD
result,  \cite{kirArnold}: \be \zeta/s \propto \log^{-2}(\pi
T/\Lambda)\log^{-1}\log(\pi T/\Lambda)\;. \ee

\subsection{Holographic explanation for the rise of $\zeta/s$ near $T_c$ and the small black-hole branch}

 With the same numerical methods, one can also compute the ratio $\zeta/s$ on the
 small black-hole\index{black hole!small}
 branch. As this solution has a smaller value of the
 action than the large black-hole solution, it is a subleading saddle point in the phase space of the theory,
 hence bears no direct significance for an holographic investigation of the
 quark-gluon plasma. However, as we show below, the existence of this branch
 provides  a holographic explanation for the peak
 in $\zeta/s$ in the quark-gluon plasma, near $T_c$.

\begin{figure}
 \begin{center}
\includegraphics[scale=0.7]{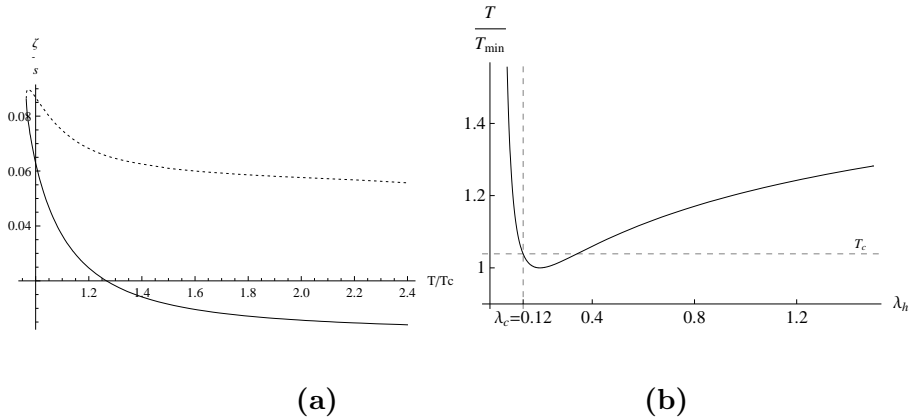}
\includegraphics[scale=0.7]{Tlh-0403.eps}
 \end{center}
 \vspace*{3pt}
 \centerline{\hspace{.3in} {\bf (a)} \hspace{1.5in}
{\bf (b)}}
 \caption{{\bf (a)}\,\, Numerical evaluation of
$\zeta/\eta$ both on the large-BH branch (the solid curve) and on
the small BH branch (the dashed curve). $T_m$ denotes $T_{min}$.
{\bf (b)} \,\, The two branches of black-hole solutions, that
correspond to different ranges of $\l_h$. The large BH corresponds
to $\l_h<\l_{min}$ and the small BH corresponds to
$\l_h>\l_{min}$.}
 \label{kirillus}
 \end{figure}

\begin{figure}
 \begin{center}
\includegraphics[height=6cm,width=9cm]{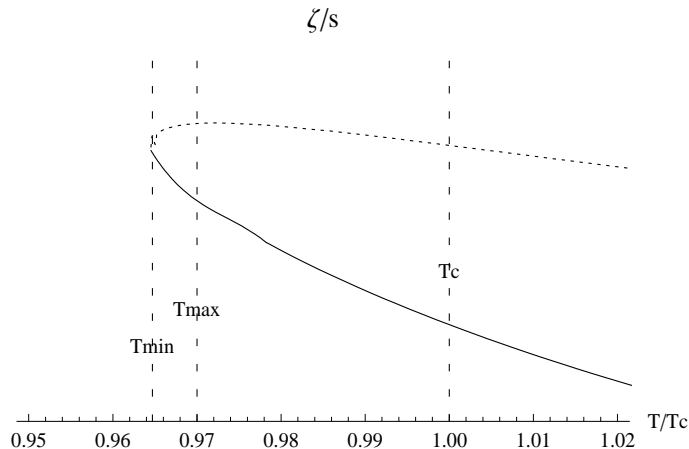}
 \end{center}
 \caption[]{An inset from the figure \ref{kirillus} around the maximum of $\zeta/s$. }
\label{kirbulkviscinset}
 \end{figure}

 From the practical point of view, we find the second numerical method above (solving the fluctuation equation at $\omega=0$)
 easier in the range of $\l_h$ that corresponds
to the small black hole. The result is shown in figure \ref{kirillus}
(a). The presence of two branches for $T>T_{min}$, is made clear
in this figure. See also fig \ref{kirillus} (b) for the respective ranges of
$\l_h$ that correspond to small and large BHs. In fig \ref{kirillus}
(a), $\zeta/s$ on the large BH branch is depicted with a solid curve
and the small BH branch is depicted with a dashed curve. We
observe that $\zeta/s$ keeps increasing on the large-BH branch as T
is lowered, up to the temperature $T_{min}$ where the small and
large BH branches merge\footnote{As far as the thermodynamics of the
gluon plasma is concerned, the temperatures below $T_c$ (on the
large BH branch) has little importance, because for $T<T_c$
the plasma is in  the confined phase.}. On the other hand, on the
small BH branch $\z/s$ keeps increasing as the T is increased, up
to a certain $T_{max}$ that lies between $T_{min} $ and $T_c$, see
figure \ref{kirbulkviscinset}. From this point onwards, $\z/s$
decreases with increasing T.

A simple fact that can be proved analytically is that the
derivative of $\zeta/s$ diverges at $T_{min}$. This is also clear
from figure \ref{kirbulkviscinset}. This is shown by inspecting
 equation (\ref{kirzs}). The T derivative is determined as
$d/dT = (dT/d\l_h )d/d\l_h$. Whereas the derivative w.r.t $\l_h$
is everywhere smooth\footnote{Note that $c_b$ is also a function
of $\l_h$. As both the fluctuation equation (\ref{kirhw}) and the
boundary conditions are smooth at $\l_h=\l_{min}$, one concludes
that $c_b$ also is smooth at this point.}, the factor $dT/d\l_h$
diverges at $T_{min}$ by definition, see figure \ref{kirillus} (b).

Therefore,  the presence of a $T_{min}$ where the
large and the small black holes meet,\index{black hole!small}\index{black hole!large} in other words, the presence of a
small black-hole branch is  responsible for the increase of $\z/s$
near $T_{min}$. As in most of the holographic constructions that
we analyzed, and specifically in the example we present here,
 $T_c$ and $T_{min}$ are close to one another, this fact
implies a rise in the bulk viscosity near $T_c$. This proposal,
combined with the fact that {\em the existence of a small BH
branch and color confinement in the dual gauge theory at zero T
are in one-to-one correspondence} \cite{kirGKMN2}, suggests that {\em
in confining large-N gauge theories, there will be a peak in the
ratio $\zeta/s$ close to $T_c$.}

Another fact that can be shown analytically is that $\zeta/s$
asymptotes to a finite value as $T\to\infty$ in the small
black-hole branch \footnote{See the discussion at appendix B of \cite{kirdrag}.}. We find that,
\be\lab{kirsbhlim} \frac{\zeta}{s}\bigg|_{small}
\to \frac{1}{6\pi}, \qquad as\,\,\, T\to\infty.
 \ee
As the entropy density vanishes in this limit \cite{kirGKMN2}, we
conclude that $\zeta$ should vanish with the same rate.

For a general potential with strong coupling asymptotics \be
V(\l)\sim \l^{Q}~~~~~{\rm as}~~~~\l\to\infty, \ee taking into
account (\ref{kirzs}), equation (\ref{kirsbhlim}) is modified to
\be\frac{\zeta}{s}\bigg|_{small} \to \frac{3Q^2}{32\pi}, \qquad
as\,\,\, r_h\to r_0. \label{kiran1} \ee where $r_0$ is the position
of the singularity in the zero temperature solution.

For confining
theories, the limit $r_h\to r_0$ corresponds to $T\to\infty$ on
the small BH branch. However, one can show that the result
(\ref{kiran1}) holds quite generally, regardless of whether the zero T theory
confines or not. In particular, for the non-confining
theories---that is either when $Q<4/3$ or when $Q=4/3$ but the
subleading term in the potential vanishes at the singularity---there
is only the large black-hole branch\index{black hole!large} and the limit $r_h\to r_0$
corresponds to the zero T limit of this BH. Thus, we also learn
that there exist holographic models that correspond to non-confining gauge theories
whose zero T limit yield a
constant $\zeta/s$. This constant approaches zero as $Q\to 0$, i.e.
in the limiting AdS case.

We also see that the asymptotic value of $\z/s$ in the small BH
branch is close to the value of $\z/s$ near $T_c$. We shall give
an explanation of this fact in the next subsection. Using the
asymptotic formula (\ref{kiran1}), the fact that $Q>{4\over 3}$ for
confinement and $Q\leq {4\sqrt{2}\over 3}$ for the IR singularity
to be good and repulsive we may obtain a range of values where we
expect $\z/s$ to vary, namely \be {1\over 6\pi}~~\leq~~
\frac{\zeta}{s}\bigg|_{\rm small, asymptotic}~~\leq~~ {1\over
3\pi}. \label{kiran2}
 \ee

\begin{figure}
 \begin{center}
\includegraphics[height=6cm,width=9cm]{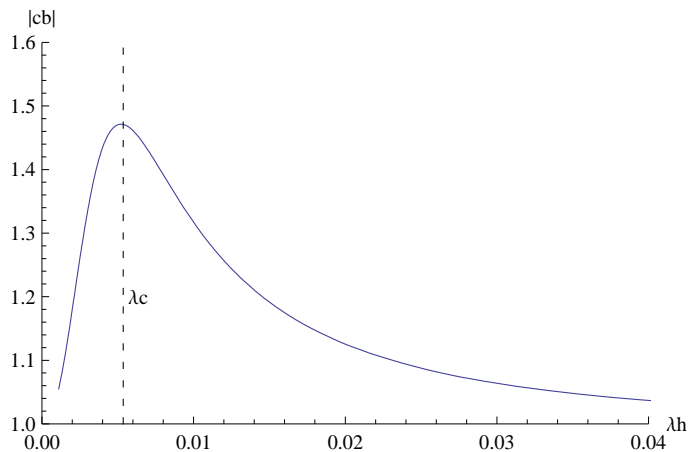}
 \end{center}
 \caption[]{The coefficient $|c_b|$ of equation (\ref{kirzs}) as a function of $\l_h$.}
\label{kircblh}
 \end{figure}

A final observation concerns the coefficient $c_b(\l_h)$ in
(\ref{kirzs}). This part is the only input from the solution of the
fluctuation equation, the rest of (\ref{kirzs}) is  fixed by the
dilaton potential entirely. We plot the numerical result for $c_b$ in fig \ref{kircblh}
as a function of the coupling at the horizon $\l_h$.

First of all, Figure \ref{kircblh} provides a
check that, the approximate bound of \cite{kirGubser:2008sz}
$|c_b|\geq1$, is satisfied in the entire range. One also observes
$c_b$ approaches to 1 in the IR and UV asymptotics. These facts can
be understood analytically: In the UV (near the boundary) it is
because of the boundary condition $c_b=1$.
In the IR, it is more
subtle, and it is explained in appendix B of \cite{kirdrag}.

Finally, we observe that the
deviation of $c_b$ from the asymptotic value 1 is maximum around
the phase transition point $\l_c$. In fact, we numerically
observed that the top of the curve in figure \ref{kircblh} coincides
with $\l_c$ to a very high accuracy. Whether this is just a
coincidence or not, remains to be clarified.

\subsection{The adiabatic approximation}

Motivated by the Chamblin-Reall solutions\index{Chamblin-Reall solution} \cite{kirCR}, Gubser et al.
\cite{kirgubser} proposed an approximate adiabatic formula for
the speed of sound.\index{speed of sound} In the case when $V'/V$ is a slowly varying
function of $\f$, \cite{kirgubser} proposes the following
formulae for the entropy density and the temperature:
\bea \log s
&=& -\frac83\int^{\f_h} d\f \frac{V}{V'} + \cdots,\lab{kiradb1}\\
\log T &=& \int^{\f_h} d\f \le({1\over 2} \frac{V'}{V}
-\frac89\frac{V}{V'}\ri) \cdots,\lab{kiradb2}
\eea
where the ellipsis
denote contributions slowly varying in $\f_h$\footnote{Various
coefficients in these equations  differ from \cite{kirgubser}
due to our different normalization of the dilaton kinetic term.}.

\begin{figure}
 \begin{center}
\includegraphics[height=6cm,width=9cm]{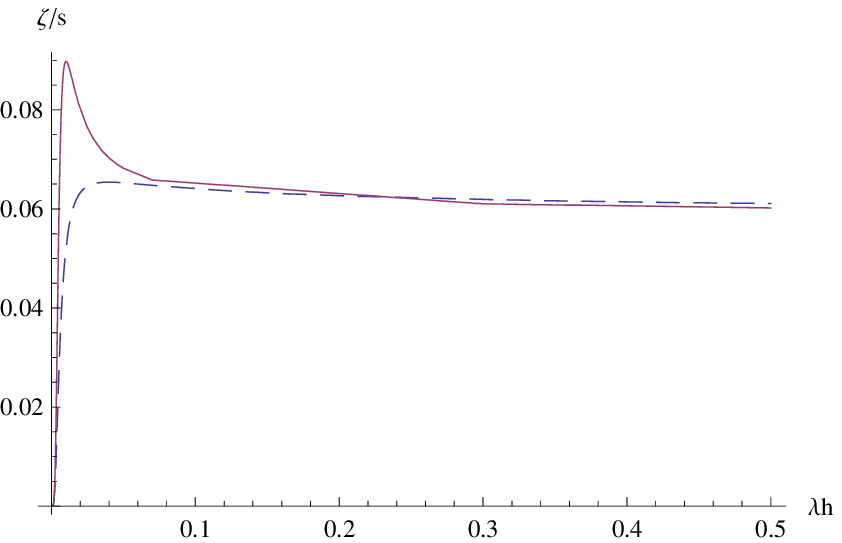}
 \end{center}
 \caption[]{Comparison of the exact $\zeta/s$ with the adiabatic approximation in the variable $\l_h$.
 Solid(red) curve is the full numerical result and the dashed(blue) curve follows from (\ref{kirzsadb}).}
\label{kirfigadb1}
 \end{figure}

It is very useful to reformulate this approximation using the
method of scalar variables, which in turn allows us to extract the
general T dependence of most of the thermodynamic observables in
an approximate form. Here, we apply this formalism to the
computation of $\zeta/s$. The method of scalar
variables and the details of the adiabatic
approximation in the scalar variables are given in \cite{kirdrag}.

For the scalar variable $X={\Phi'\over 3A'}$ the adiabatic approximation means
\be\lab{kirXadb}
X(\f) \approx -\frac38 \frac{V'(\f)}{V(\f)}.
\ee
 The fluctuation equation
(\ref{kirhw}) greatly simplifies with (\ref{kirXadb}). In fact, as shown
in \cite{kirdrag}, the solution becomes independent of $\f$. With unit
normalization on boundary, the adiabatic solution in the entire
range of $\f\in(-\infty, \f_h)$ becomes $h_{adb}(\f)=1$.
Consequently, the coefficient $c_b$ in (\ref{kirzs}) becomes unity,
hence:
\be
 \label{kirzsadb} \frac{\zeta}{s}\bigg|_{adb} =
\frac{3}{32\pi} \left(\frac{V'(\f_h)}{V(\f_h)}\right)^2.
\ee

\begin{figure}
 \begin{center}
\includegraphics[height=6cm,width=9cm]{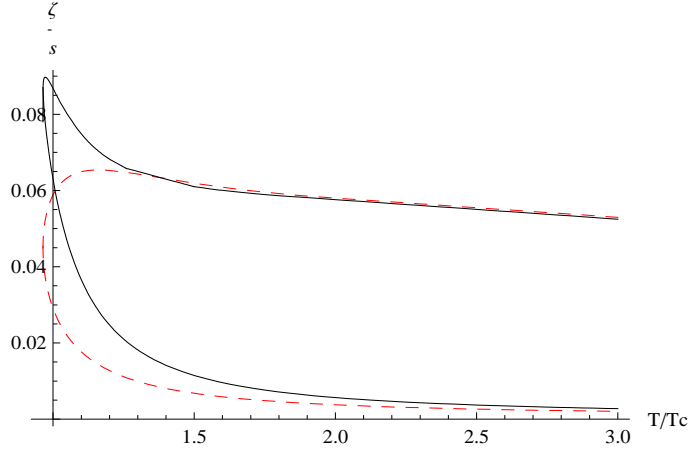}
 \end{center}
 \caption[]{Comparison of the exact $\zeta/s$ with the adiabatic approximation in variable T. Solid(blue) curve is the full numerical
 result and the dashed(red) curve follows from (\ref{kirzsadb}).}
\label{kirfigadb2}
 \end{figure}

We
plot this function in $\l_h$ in figure \ref{kirfigadb1}, where we also
provide the exact numerical result for comparison. Note that in figure \ref{kirfigadb1}
the whole large black-hole\index{black hole!large} branch has been compressed at the left of the figure for $\l_h\lesssim 0.04$
The same
functions in the variable $T/T_c$ are plotted in figure \ref{kirfigadb2}.

The validity of the adiabatic approximation equation (\ref{kirXadb}), is
determined by the rate at which $V'/V$ varies with $\Phi$.  In
particular, the approximation becomes exact in the limits where
$V'/V$ becomes constant. This happens for a constant potential or
 a potential that is a single power of $\l$ (exponential in $\Phi$).
This is the case in the UV
($\f\to-\infty$, where the potential becomes a constant) and the
IR ($\f\to + \infty$ where the potential becomes a power law.)
. Therefore
equation (\ref{kirzsadb}) allows us to extract the analytic behavior of
$\zeta/s$ in the limits $\Phi_h\to\pm\infty$.

 The numerical values
one obtains from (\ref{kirzsadb}) in the intermediate region may
differ from the exact result (\ref{kirzs}) considerably, especially
near $T_c$. However, we expect that the general shape will be
similar.

Finally, the adiabatic approximation hints at why, in the
particular background that we study, $\zeta/s$ at $T_c$ is close
to the limit value (\ref{kirsbhlim}): In order to see this we rewrite (\ref{kirzsadb})
as \be \label{kirzsadb2} \frac{\zeta}{s}\bigg|_{adb} =
\frac{2}{3\pi} X^2. \ee In the limit (\ref{kirsbhlim}) we have
 $X\to -1/2$. The only other point where $X = -1/2$, is at the
 minimum of the string frame scale factor $\f_*$. This is the
 point where the confining string saturates \cite{kirihqcd}.
On the other hand, we expect on general physical grounds that the
de-confinement phase transition happens near this point, i.e.
$\f_c\approx \f_*$. Thus, the adiabatic formula predicts that
$\zeta/s (\f_c)$ be close to the limit value
$1/6\pi$.\footnote{This argument may break down for two (dependent) reasons:
First of all the adiabatic approximation becomes lees good
near $\f_c$. This is because, in this region $V'/V$ varies
relatively more rapidly as a function of $\f$. Secondly, precisely
because of this, even though $\f_c$ is not far away from $\f_*$
the difference can result in a considerable change in the value of
$\zeta/s$ through (\ref{kirzsadb}).}

\begin{figure}[h!]
 \begin{center}
\includegraphics[height=4cm,width=7cm]{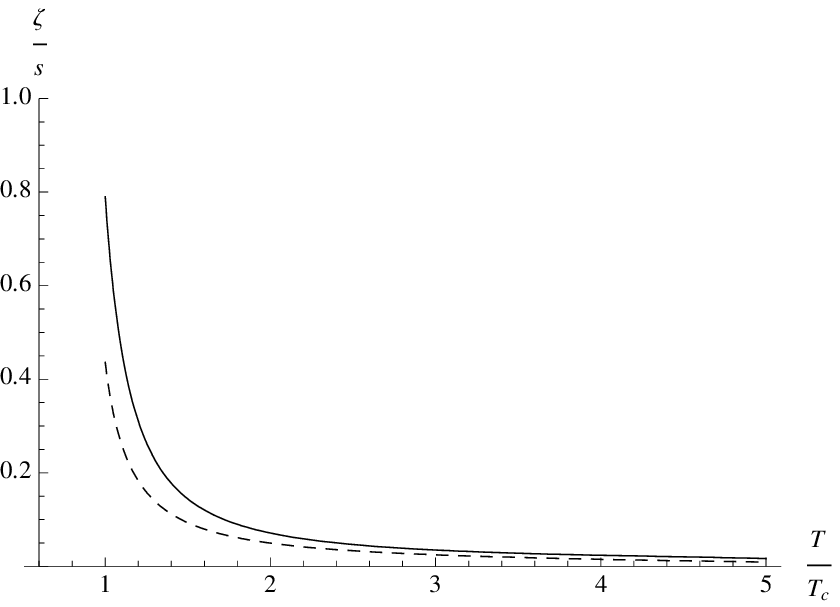}\\
(a)\\
\includegraphics[height=4cm,width=7cm]{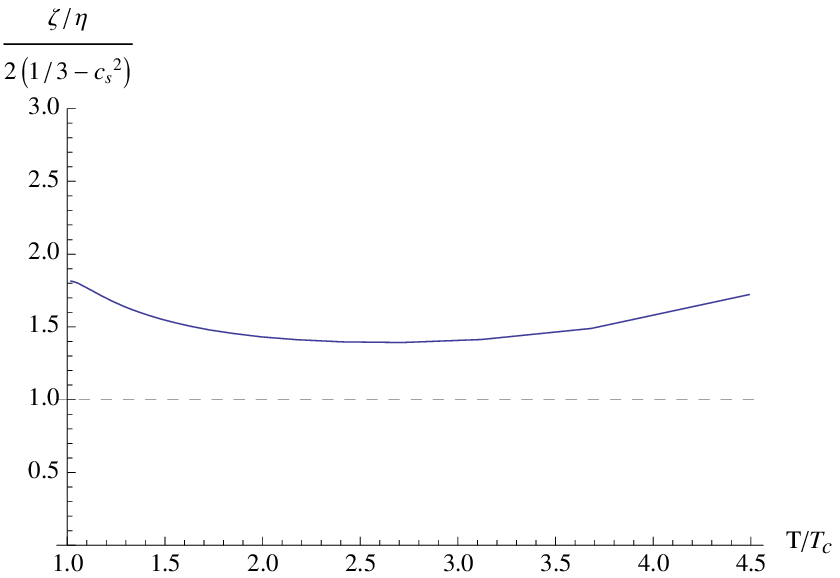}\\
(b)
 \end{center}
 \caption[]{(a) Comparison of $\zeta/\eta$ (solid line) and the RHS of (\ref{kirbuchel1}) (dashed line),
obtained using the speed of sound of the IHQCD model \cite{kirfit}.
(b) Plot of  the function $C(T)$ defined in equation (\ref{kirbuchel2}) as a
function of temperature. The horizontal dashed line indicates where
Buchel's bound is saturated.
 We see that the bound is satisfied in the entire range of temperatures.}
\label{kirbuchelfig}
 \end{figure}

\subsection{Buchel's bound}

In \cite{kirBuchel}, Buchel proposed a bound  for the ratio of the bulk and shear viscosities,
 motivated by certain well-understood holographic examples.
In 4 space-time dimensions the Buchel bound\index{Buchel bound} reads,
\be\label{kirbuchel1}
\frac{\zeta}{\eta} \geq 2\left(\frac13 -c_s^2\right).
\ee
We note that the bound is proposed to hold in the entire range of temperature from $T_c$ to $\infty$.
This bound is trivially satisfied for exact conformal theories such as ${\cal N}=4$ YM, and
saturated in theories on $Dp$ branes \cite{kirBuchel,kirkas}.
With the numerical evaluation at hand, we can check (\ref{kirbuchel1}) in our case. In figure \ref{kirbuchelfig} (a)  we plot the LHS and RHS of the bound
\footnote{Since this theory contains two derivatives only, ${\eta\over s}$ has the
 universal value $1/4\pi$.}. We clearly
see that the bound is satisfied for all temperatures. As expected, both the LHS and the RHS of (\ref{kirbuchel1}) vanishes in the high T conformal limit.

A clear picture of Buchel's bound is obtained by defining the function:
\be\label{kirbuchel2}
C(T) = {\zeta/\eta \over 2\left(1/3 -c_s^2\right)},
\ee
in terms of which the bound is simply $C > 1$. In Figure \ref{kirbuchelfig} (b)
 we show the function $C(T)$  obtained numerically in our  IHQCD model, between $T_c$  and  $5T_c$.
  The values of this function are mildly dependent
on temperature, and are between 1.5 and 2, the same range of values that were recently considered in the hydrodynamic codes
by Heinz and Song \cite{kirHeinz}.

We may also investigate the fate of the bound at large T.
using the asymptotics of $\zeta/s$ in (\ref{kirhighTbBH})
\be \lab{kirhighTbBH1}
\frac{\z}{s}\bigg|_{\rm large}= \frac{2}{27\pi}\frac{1}{\log^2(\pi
T/\Lambda)}+\cdots, \qquad as\,\,\, T\to\infty. \ee
and
\be\lab{kircshT}
\frac{1}{c_s^2}-3 = \frac{4}{3}\frac{1}{\log^2\le(\frac{T}{T_c}\ri)}
+\frac{32b}{9} \frac{\log\le(\log\le(\frac{T}{T_c}\ri)\ri)}{\log^3\le(\frac{T}{T_c}\ri)}+\cdots
\ee
from \cite{kirGKMN2}, that can be rewritten as
\be
{1\over 3}-c_s^2=\frac{4}{27}\frac{1}{\log^2\le(\frac{T}{T_c}\ri)}
+\frac{32b}{81} \frac{\log\le(\log\le(\frac{T}{T_c}\ri)\ri)}{\log^3\le(\frac{T}{T_c}\ri)}+\cdots
\label{kirvs2}\ee
where $b={b_1\over b_0^2}={3\cdot 34\over 2\cdot 121}$ is the ratio of the two-loop to the one-loop squared
$\beta$-function coefficients in large-$N_c$ YM.

Since  in this class of models $\eta/s=1/4\pi$ exactly we obtain
\be
\lim_{T\to\infty}{\zeta/\eta \over 2\left(1/3 -c_s^2\right)}=1
\ee
in agreement with a recently derived general formula, in Einstein dilaton gravity, \cite{kiryarom},
\be
\lim_{T\to\infty}{\zeta/\eta \over 2\left(1/3 -c_s^2\right)}=2\pi{4-\Delta\over 4-2\Delta}\cot\left({\pi \Delta\over 4}\right)
\ee
where $\Delta$ is the scaling dimensions of the scalar operator in the UV, that is marginal in our case.

It has also been suggested recently, \cite{kircherman,kirstephanov}, that the speed of sound squared, in Einstein dilaton gravity
asymptotes to 1/3 at high temperatures from below. This is evident in our asymptotic formula  (\ref{kirvs2}), although
the formulae in \cite{kircherman,kirstephanov} fail to capture correctly the marginal case that is relevant here.

\section{The drag force on strings and heavy quarks}

We will now consider an (external) heavy quark\index{heavy quark} moving through an infinite volume of
gluon plasma with a fixed velocity $v$ at a finite temperature T
\cite{kirher,kirgub}.
  The quark feels a drag force\index{drag force} coming from its interaction with the plasma
  and an external force has to be applied in order for it to keep a constant velocity.
  In a more realistic set up one would like to describe the deceleration caused by the drag.

  The heavy external quark can be described by a string whose endpoint is at the boundary.
  One can accommodate flavor by introducing D-branes, but we will not do this here.
  A first step is to describe the classical string ``trailing" the quark.

  We consider the Nambu-Goto action on the world-sheet of the string.
\begin{equation}\label{kirNGACTION}
S_{NG} = -\frac{1}{2\pi \ls^2}\int d\sigma d\tau \sqrt{det\left(-g_{MN} \partial_{\alpha}X^{M} \partial_{\beta}X^{N}\right)} \;,
\end{equation}
where the metric is the string frame metric.
The ansatz we are going to use to describe the trailing string is, \cite{kirgub},
\begin{equation}
\label{kirTRAILANSATZ}
 X^{1} = v t +\xi(r),\quad X^{2}=X^{3}=0 \;,
\end{equation}
along with the gauge choice
\begin{equation}\label{kirTRAILGAUGE}
\sigma =r, \quad \tau =t\;,
\end{equation}
where $r$ is the (radial) holographic coordinate. The string is moving in the $X^1$ direction.

This is a ``steady-state" description of the moving quark as acceleration and deceleration are not taken into account.
  For a generic background the action of the string becomes
\begin{equation}
S = -\frac{1}{2\pi \ls^2} \int dt dr ~\sqrt{-g_{00}g_{rr}-g_{00}g_{11}\xi'^{2}-g_{11}g_{rr}v^2} \;.
\end{equation}
Note that $g_{00}$ is negative, and we should check whether our solution produces a real action.
For example a straight string stretching from the quark to the horizon is a solution to the equations of
 motion but has imaginary action.

 We note that the action does not depend on $\xi$ but only its derivative, therefore the corresponding ``momentum" is conserved
\begin{equation}\label{kirCONSMOMENTUM}
\pi_{\xi} = -\frac{1}{2\pi \ls^2} \frac{g_{00}g_{11}\xi'}{\sqrt{-g_{00}g_{rr}-g_{00}g_{11}\xi'^2-g_{11}g_{rr}v^2}}\;.
\end{equation}
We solve  for $\xi'$ to obtain
\begin{equation}\label{kirGOTMOMENTUM}
\xi' = \frac{\sqrt{-g_{00}g_{rr}-g_{11}g_{rr}v^2}}{\sqrt{g_{00}g_{11}\left(1+g_{00}g_{11}/(2\pi\ls^2 \pi_{\xi})^2\right)}} \;.
\end{equation}
The numerator changes sign at some finite value of the fifth coordinate $r_{s}$.
 For the solution to be real, the denominator has to change sign at
  the same point.  We therefore determine $r_{s}$ via the equation
\begin{equation}\label{kirGOTZS}
 g_{00}(r_{s})+g_{11}(r_{s})v^2=0\;,
\end{equation}
and the constant momentum
\begin{equation}\label{kirGOTPIXI}
 \pi_{\xi}^{2} = -\frac{g_{00}(r_{s})g_{11}(r_{s})}{(2\pi \ls^2)^2}\;.
\end{equation}
Writing the string-frame metric as
\be
ds^2=e^{2A_s}\left[{dr^2\over f}-f~dt^2+dx\cdot dx\right]
\label{kirb1}\ee
(\ref{kirGOTZS}) becomes
\be
v^2=f(r_s)
\label{kirb3}\ee
The induced world-sheet metric is therefore
\be
g_{\a\b} = e^{2A_s(r)}\left( \begin{array}{cc}
-(f(r)-v^2)~~ & ~~{e^{2A_s(r_s)}v^2\over f(r)}\sqrt{f(r)-v^2\over  e^{4A_s(r)}f(r)-e^{4A_s(r_s)}v^2} \\
{e^{2A_s(r_s)}v^2\over f(r)}\sqrt{f(r)-v^2\over  e^{4A_s(r)}f(r)-e^{4A_s(r_s)}v^2}~~ & {e^{4A_s(r)}f^2(r)-v^4e^{4A_s(r_s)}\over
f^2(r)\left[e^{4A_s(r)}f(r)-v^2{e^{4A_s(r_s)}}\right]}
\end{array}\right)\label{kirindu}
\end{equation}

We can change the time coordinate to obtain a diagonal induced metric $t=\tau+\zeta(r)$ with
$$
\zeta'={e^{2A_s(r_s)}v^2\over f(r)\sqrt{(f(r)-v^2)( e^{4A_s(r)}f(r)-e^{4A_s(r_s)}v^2)}}
$$
The new metric is
\be
ds^2=e^{2A_s(r)}\left[-(f(r)-v^2)d\tau^2+{e^{4A_s(r)}\over (e^{4A_s(r)}f(r)-e^{4A_s(r_s)}v^2)}dr^2\right]
\end{equation}
and near $r=r_s$ it has the expansion
\be
ds^2=\left[-f'(r_s)e^{2A_s(r_s)}(r-r_s)+{\cal O}((r-r_s)^2)\right]d\tau^2+\left[{e^{2A_s(r_s)}\over
(4v^2 A'_s(r_s)+f'(r_s))(r-r_s)}+{\cal O}(1)\right]dr^2
\ee
This is a world-sheet black-hole\index{black hole!world-sheet} metric with horizon at the turning point $r=r_s$.

\subsection{The drag force}

The drag force\index{drag force} on the quark can be determined by calculating  the momentum that is lost by flowing along  the string into the horizon:
\begin{equation}\label{kirGOTDRAGFORCE}
F_{\rm drag} = \frac{dp_{1}}{dt}=-\frac{1}{2\pi\ls^2}\frac{g_{00}g_{11}\xi'}{\sqrt{-g}}=\pi_{\xi}\;.
\end{equation}
 This can be obtained by considering the world-sheet Noether currents\index{Noether current} $\Pi^{\alpha}_{M}$
and expressing the loss of momentum  as $\Delta P^{z}_{x_1} = \int\Pi^{r}_{1} $.
This may be  evaluated  at any
   value of r, but it is more convenient to evaluate it at $r=r_{s}$.

   We finally find that
\begin{equation}\label{kirDRAGEXPL}
 F_{\rm drag} = - \frac{1}{2\pi \ls^2} \sqrt{-g_{00}(r_{s})g_{11}(r_{s})}\;.
\end{equation}

Using the form (\ref{kirb1}) of our finite-temperature metric in the string frame
we finally obtain
\be
 F_{\rm drag} = - \frac{e^{2A_s(r_s)}\sqrt{f(r_s)}}{2\pi \ls^2}=-\frac{ e^{2A(r_{s})}\sqrt{f(r_s)} \lambda(r_{s})^{4/3}}{2\pi\ls^2}\;,
\label{kirb2}\ee
where in the second equality we expressed the force
in terms of the Einstein-frame scale factor and the ``running" dilaton.
Substituting from (\ref{kirb3}) we obtain
\be
 F_{\rm drag} = - \frac{v~e^{2A_s(r_s)}}{2\pi \ls^2}=-\frac{ v ~e^{2A(r_{s})}\lambda(r_{s})^{4/3}}{2\pi\ls^2}\;,
\label{kirb4}\ee
Before proceeding further, we will evaluate the drag force for the conformal case of ${\cal N}=4$ SYM
where
\be
e^{A_s}={\ell\over r}\sp v^2=f(r_s)=1-(\pi T r_s)^4\sp {\ell^2\over \ls^2}=\sqrt{\l}
\label{kirb5}\ee
Substituting in (\ref{kirb4}) we obtain, \cite{kirher}-\cite{kirgub},
\be
F_{\rm conf}={\pi\over 2}\sqrt{\lambda}~T^2{v\over \sqrt{1-v^2}}
\label{kira9}\ee

Moving on to YM, to compute the drag force from equation (\ref{kirb4}) we must
first determine $\ls$ in the IHQCD model. In this setup there is no analog
of the ${\cal N}=4$ SYM relation (\ref{kirb5}) between $\ell$,  $\ell_s$ and $\l$.
Rather, the fundamental string length $\ls$ is determined in a bottom-up fashion,
 by matching the effective string tension to the QCD string tension $\sigma_c$
derived from the lattice calculations.  We obtain
\begin{equation}
\sigma_c= \frac{1}{2\pi\ls^2} e^{2A_{s,o}(r_{*})} =\frac{1}{2\pi\ls^2} e^{2A_o(r_{*})} \lambda_o(r_{*})^{4/3}\;,
\label{kirb6}\end{equation}
where $r_{*}$ is the point where the zero-temperature string scale  factor (at T=0)  $A_{s,o} (r)$ has a minimum.
 For a typical value of $\sigma_c\sim (440 \;MeV)^2$ \cite{kirchenetal} we find
\begin{equation}
 \ell_{s} = 6.4 ~\ell \;,
\label{kirb7}\end{equation} where $\ell$ is the radius of the
asymptotic AdS space.

 On the other hand, unlike in ${\cal N} = 4$ SYM,
in the IHQCD model the value of the coupling $\l(r_s)$ in equation (\ref{kirb4})
is not an extra parameter to be fixed by hand, but rather it is determined dynamically together with the background metric.

\begin{figure}[ht]
\centering
\includegraphics[width=10cm]{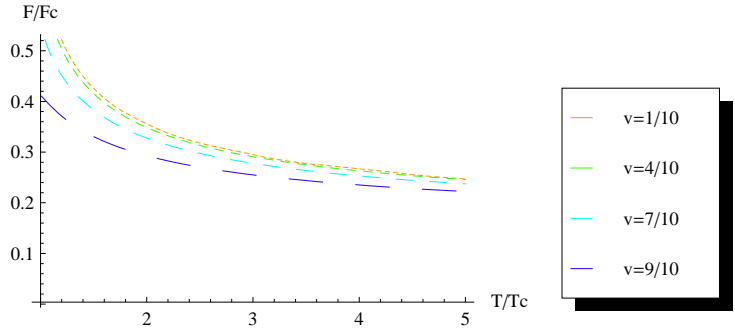}
\caption{In this figure the ratio of the drag force in improved
holographic QCD to the drag force in ${\cal N}=4$ SYM is shown.
The ratio is computed for different velocities as a function of
temperature. The 't Hooft coupling for the ${\cal N}=4$ SYM
theory is taken to be $5.5$. We chose this value as it is considered
in the central region of possible values for the 't Hooft coupling. It is seen that as
the velocity increases the value of the ratio decreases.}\label{kirMyFdragFigureA}
\end{figure}

\begin{figure}[ht]
\centering
\includegraphics[width=10cm]{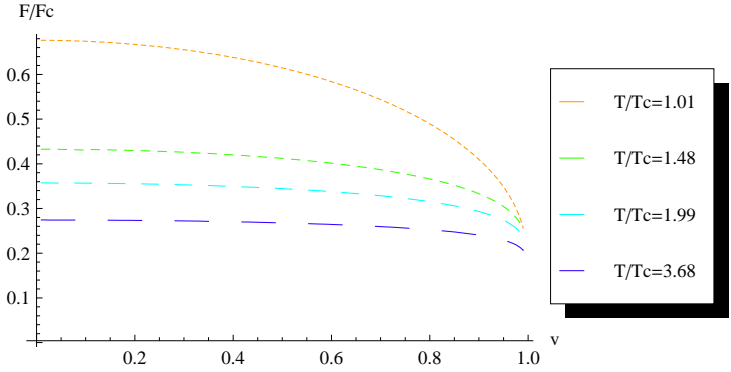}
\caption{In this figure the ratio of the drag force in improved holographic QCD to the drag force in ${\cal N}=4$ SYM is shown.
The ratio is computed for different temperatures as a function of velocity.
The 't Hooft coupling for the ${\cal N}=4$ SYM theory is taken to be $5.5$.
As temperature increases the value of the ratio decreases.}\label{kirMyFdragFigureB}
\end{figure}

\subsection{The relativistic asymptotics}

When $v\to 1$ then $r_s\to 0$ and we approach the boundary.
Near the boundary ($r\to 0$) we have the following asymptotics of the scale factor and the 't Hooft coupling, \cite{kirihqcd}
\be
f(r)\simeq 1-{\pi T~e^{3A(r_h)}\over \ell^3}r^4\left[1+{\cal O}\left({1\over \log(\Lambda r)}\right)\right]+{\cal O}(r^8)
\sp e^{A(r)}={\ell\over r}\left[1+{\cal O}\left({1\over \log(\Lambda r)}\right)\right]+\cdots
\label{kira2}\ee
and
\be
\lambda(r)=-{1\over \beta_0\log(r\Lambda)}+{\cal O}(\log(r\Lambda)^{-2})
\label{kirb8}\ee
where $r_h$ is the position of the horizon.

We therefore obtain for the turning point
\be
r_s\simeq \left[{\ell^3(1-v^2)\over \pi T e^{3A(r_h)}}\right]^{1\over 4}\left[1+{\cal O}\left({1\over \log(1-v^2)}\right)\right]
\sp \lambda(r_s)\simeq
-{4\over \beta_0 \log \left[1-v^2\right]} +\cdots
\label{kira3}\ee
and the drag force
\be
F_{\rm drag}\simeq -{\sqrt{\pi T\ell b^3(r_h)\lambda^{8\over 3}(r_s)}\over 2\pi\ell_s^2}{v\over \sqrt{1-v^2}}+\cdots
\label{kira4}\ee
We also use
\be
e^{3A(r_h)}={s(T)\over 4\pi M_p^3~ N_c^2}={45\pi\ell^3 s(T)\over N_c^2}
\label{kira77}\ee
where $s(T)$ the entropy per unit three-volume, and we write the relativistic asymptotics of the drag force as,
\be
F_{\rm drag}\simeq -{\sqrt{\pi T\ell b^3(r_h)}\over 2\pi\ell_s^2}{v\over \sqrt{1-v^2}\left(-{\beta_0\over 4}\log\left[1-v^2\right]\right)^{4\over 3}}+\cdots
\label{kirb9}\ee
$$
=-{\ell^2\over \ell_s^2} \sqrt{{45 ~T s(T)}\over 4N_c^2}{v\over \sqrt{1-v^2}\left(-{\beta_0\over 4}\log\left[1-v^2\right]\right)^{4\over 3}}+\cdots
$$

The force is proportional to the relativistic  momentum
combination $v/\sqrt{1-v^2}$ modulo a power of
$\log\left[1-v^2\right]$. This factor is present because, as
argued in \cite{kirdiss} the asymptotic metric is AdS in the Einstein
frame instead of the string frame. Its effects are not important
phenomenologically.

\subsection{The non-relativistic asymptotics}

We now consider the opposite limit,  $v\to 0$. In this case the turning point asymptotes to the horizon, $r_s\to r_h$
and we have the expansion
\be
f(r)\simeq 4\pi T(r_h-r)+{\cal O}((r_h-r)^2)\sp r_s=r_h-{v^2\over 4\pi T}+{\cal O}(v^4)
\label{kira10}\ee
and
\be
F_{\rm drag}\simeq -{e^{2A(r_h)}\lambda(r_h)^{4\over 3}\over 2\pi\ell_s^2}v\left[1-{v^2\over 2\pi T}A'(r_h)-
{v^2\over 3\pi T}{ \lambda'(r_h)\over \lambda(r_h)}+{\cal O}(v^4)\right]
\label{kira11}\ee
$$
\simeq -{\ell^2\over \ls^2}\left({45\pi ~s(T)\over N_c^2}\right)^{2\over 3}{\l(r_h)^{4\over 3}\over 2\pi}v+{\cal O}(v^3)
$$
where primes are derivatives with respect to the conformal coordinate $r$.

\subsection{The diffusion time}

For a  heavy quark with mass $M_q$ we may rewrite
(\ref{kira9}) as
\be
F_{\rm conf}\equiv {dp\over dt}=-{1\over \tau}p\sp p={M_q v\over \sqrt{1-v^2}}
\label{kirb10}\ee
where the first equation defines the diffusion time $\tau$. In the conformal case, the diffusion time
is constant,
\be\label{kirb10-2}
\tau_{conf}={2M_q\over \pi \sqrt{\l}~T^2}
\ee
This is not anymore the case in QCD, where $\tau$ defined as above is momentum dependent.
We may still define it as in (\ref{kirb10}) in which case we obtain the following limits
\be
\lim_{p\to\infty}~\tau=M_q~{\ls^2\over \ell^2} \sqrt{4N_c^2\over 45 ~T s(T)}
\left({\beta_0\over 4}\log{p^2\over M_q^2}\right)^{4\over 3}+\cdots
\label{kirb11}\ee
\be
\lim_{p\to 0}~\tau=M_q~{\ls^2\over \ell^2}\left({N_c^2\over 45\pi ~s(T) }\right)^{2\over 3}{2\pi\over \l(r_h)^{4\over 3}}+\cdots
\label{kirb12}\ee

\begin{figure}[h!]
\centering
\includegraphics[width=13cm]{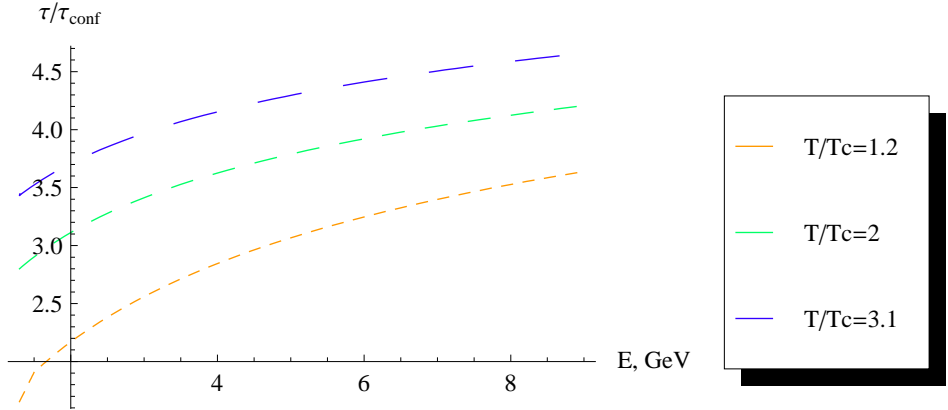}
\caption[]{
In this figure the ratio of the diffusion time in the Improved Holographic QCD model to the diffusion time in ${\cal N}=4$ SYM is shown.
 The 't Hooft coupling for ${\cal N}=4$ SYM is taken to be $\lambda=5.5$.
 The heavy quark has a mass of $M_{q}=1.3 GeV$.
  Note that with the definition of the diffusion time in (\ref{kirb10}) the ratio is the inverse of the ratio of the forces.
   A similar plot is valid for the bottom quark as well, as the mass drops out of the ratio.
     although the energy scales are different.
      In this plot the x-axis is taken to be in MeV units. As temperature increases the ratio also increases.
}\label{kirTDiffusionA}
\end{figure}

\subsection{Including the medium-induced  correction to the quark mass\label{kirmass}}

In order to estimate the diffusion time of a quark of finite  rest
mass, we must take into account the fact that the mass of the
quark receives medium-induced  corrections. In other words, the mass
appearing in equation (\ref{kirb10}) is a temperature-dependent quantity,
$M_q(T) \neq M_q(T=0)$. The ratio $M_q(T)/M_q(0)$ can be estimated
holographically by representing a static quark of finite mass by a
static, straight string \footnote{This representation ignores the
fact that the {\em kinetic} mass of a moving quark may be
different from the static mass \cite{kirher}.} stretched along the radial direction starting at a
point $r=r_q \neq 0$. At zero temperature, the IR endpoint of the
string can be taken as the ``confinement'' radius, $r_*$, where
the string frame metric reaches its minimum value; At finite
temperature, the string ends in the IR  at the BH
horizon\footnote{ It would stop at the confinement radius if the
latter were closer to the boundary than the horizon, i.e. if
$r_*(T) < r_h(T)$. However, in the model we are considering, in
the large BH branch we find that the relation $r_h < r_* $  is
always satisfied.}. The masses of the quark at zero and finite $T$
are related to the world-sheet action evaluated on the static
solution $(\tau = t, \sigma = r)$ :

\be \label{kirqmass}
M_q(0)  = {\ell \over 2\pi \ell_s^2} \int_{r_q}^{r_*} d r\, e^{2A_o(r)} \l_o^{4/3}(r)\, , \qquad   M_q(T)  = {\ell\over 2\pi \ell_s^2}  \int_{r_q}^{r_h} dr \,  e^{2A(r)} \l^{4/3}(r)\, .
\ee

\begin{figure}[ht]
\centering
\includegraphics[scale=1.2]{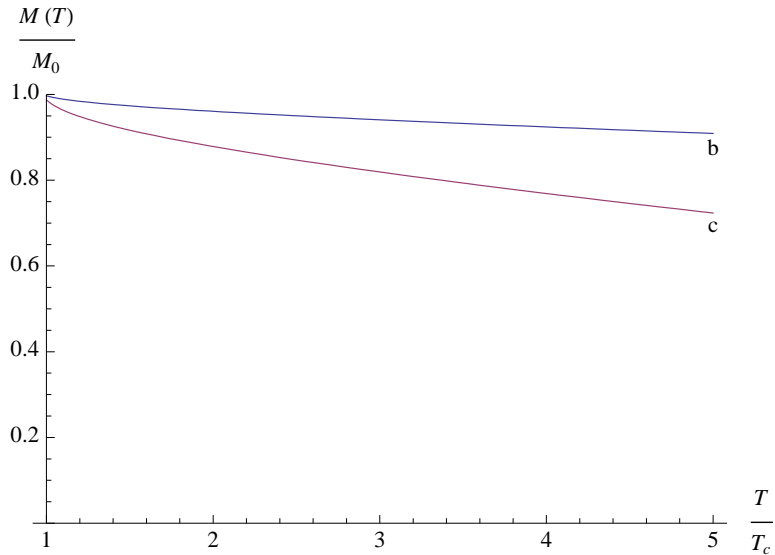}
\caption{Ratios between the thermal mass and the rest mass of the Charm (curve labeled
``c'') and Bottom (curve labeled ``b'' ) quarks, as a function of temperature.}\label{kirquarkmass}
\end{figure}

The value $r_q$ can be fixed numerically  by  matching $M_q(0)$ to the physical quark mass, and translating
the fundamental string tension  in physical units by using the relation (\ref{kirb6}),
with $\sigma_{c} = (440 MeV)^2$. This makes $M_q(T)$ a function of $M_q(0)$. The ratios $M_q(T)/M_q(0)$ we found numerically in the model
under consideration  is shown in figure \ref{kirquarkmass} for the Charm ($M(0)=1.5 GeV$)
  and Bottom ($M(0)=4.5 GeV$) quarks. The fact that, in
the deconfined plasma, the quark mass decreases with increasing temperature
is a direct consequence of the holographic framework\footnote{For a possible
field theoretical explanation  of  this phenomenon, see \cite{kirmarquet}.},
 since for higher temperature, the distance to the horizon is smaller.
An indication  that this result may be  in the right direction
comes from the lattice computation of   the
shift in  the position of the quarkonium\index{quarkonium}  resonance peak at finite temperature \cite{kirDatta}:
 in the deconfined phase the charmonium\index{charmonium}
 peak moves to lower mass at higher temperature. Our result for the
medium-induced shift in the constituent quark mass\index{constituent quark mass}  is consistent with these
observations.

\begin{figure}[ht]
\centering
\includegraphics[scale=0.8]{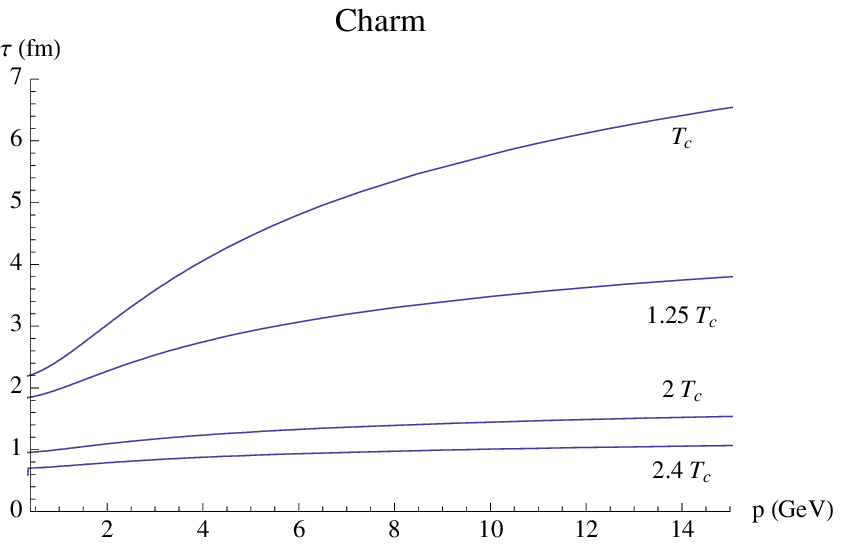}
\includegraphics[scale=0.8]{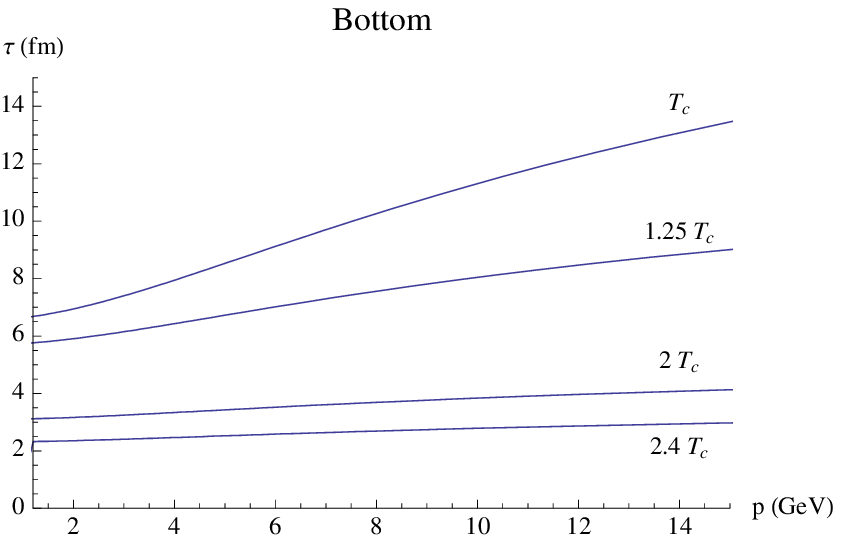}
\caption{Diffusion time for the Charm and Bottom quarks, as a function
of energy, for different ratios of the temperature to the IHQCD transition temperature $T_c$.  }\label{kirstoppingF}
\end{figure}

We can now  write the diffusion time from eqs. (\ref{kirb4}) and (\ref{kirb10}) as:
\be\label{kirstopt}
\tau(T,v) = {M_q(T) \over \sigma_{c} \sqrt{1-v^2}} \left({\l_o(r_*) \over \l(r_s)} \right)^{4/3}  e^{2A_o(r_*) - 2A(r_s)},
\ee
where once again we have eliminated the fundamental string length\index{string length}
 using equation (\ref{kirb6}). Given a set of zero- and
finite-temperature solutions, equation (\ref{kirstopt}) can be evaluated numerically
for different values of the velocity and different quark masses. The results for the
Charm ($M_q(0)=1.5 \; GeV$) and Bottom ($M=4.5 \;GeV$) quarks are displayed in figure \ref{kirstoppingF}.

\subsection{Temperature matching and diffusion time estimates}
\index{diffusion time}

An important question is how we should choose the  temperature in our holographic model in
order to compare our results with heavy-ion collision experiments. This is nontrivial, since
our setup is designed to describe pure $SU(N_c)$ Yang-Mills, whereas at RHIC temperatures
there are 3 light quark flavors that become relevant. As a consequence, the critical temperatures
and the number of degrees of freedom of the two theories are not  the same:
for pure $SU(N_c)$ Yang Mills we have $N_c^2 - 1$ degrees of freedom and a critical temperature around $260$ $MeV$; For $SU(N_c)$  QCD
with $N_f$ flavors the number of degrees of freedom is $ N_c^2 - 1  + N_c N_f$,
and the transition temperature is lower, around  $180$ MeV.

In IHQCD, the transition temperature\index{critical temperature} in physical units was calculated to be
$T_c =247$ MeV \cite{kirfit}, i.e. close to the lattice result for the pure YM  deconfining temperature. From now on, this is the
value  we will mean  when  we refer to $T_c$. This is also close to the
temperature of   QGP at RHIC,\index{RHIC} which we will denote $T_{QGP}$, and  is estimated to be  around $250$ $MeV$.
Since this value is uncertain,  below we  give our results for a range of
temperatures between 200 MeV and 400 MeV. The higher temperatures
will be relevant for the LHC\index{LHC} heavy-ion collision experiments (see e.g. \cite{kirlr3}).

Based on these considerations,  there are  different ways of fixing the temperature  (see  e.g.
the recent review \cite{kirgubser-review}): one {\em direct}  and
two {\em alternative} schemes (that we call the  {\em energy}  and {\em
entropy} scheme).
\begin{itemize}
\item {\bf{\em Direct scheme}}: The temperature of the holographic
model  is
identified with the temperature of the QGP in the experimental situation
(at RHIC or LHC),  $T_{ihqcd}^{(dir)} = T_{QGP}$.

\item {\bf{\em Energy scheme}}: One matches the energy densities, rather than the temperatures. The energy
density at RHIC is approximately (treating the QCD plasma as a free gas\footnote{This is itself
an approximation, since as we know both from experiment and in our holographic model,  the plasma is strongly
coupled up to  temperatures of a few $T_c$}.)  $\epsilon_{QGP} \simeq (\pi^2/15) (N_c^2 - 1 + N_c N_f) (T_{QGP})^4$. For $N_c=N_f=3$, asking that our
energy density matches this value requires us to consider the holographic model at temperature $T_{ihqcd}^{(\epsilon)}$ given by
\be\label{kiraltscheme}
\epsilon_{ihqcd} (T_{ihqcd}^{(\epsilon)}) \simeq 11.2 (T_{QGP})^4 
\ee
\item {\bf{\em Entropy scheme}}: Instead of matching the
energy densities, alternatively one can match the entropy density $s$,
which for the QGP, in the free gas approximation, is given   by
 $\s_{QGP} \simeq 4 \pi^2/45 (N_c^2 - 1 + N_c N_f) (T_{QGP})^4$. This leads
to the identification:
\be\label{kirentscheme}
s_{ihqcd} (T_{ihqcd}^{(s)}) = 14.9 (T_{QGP})^3
\ee
\end{itemize}

\begin{table}[h!]
\centering
\begin{tabular}{||c|c||c|c||c|c||}
\hline
$T_{QGP}$ (MeV) & $T_{QGP}/T_c$ & $T_{ihqcd}^{(\epsilon)}$ (MeV) &
$ T_{ihqcd}^{(\epsilon)}/T_c$  & $T_{ihqcd}^{(s)}$ (MeV) & $ T_{ihqcd}^{(s)}/T_c$ \\
\hline
190 & 0.77 &259 & 1.05  & 274& 1.11  \\
\hline
220 & 0.89 &290  & 1.18 & 302 & 1.23  \\
\hline
250 & 1.01 &325 & 1.31 & 335  & 1.35\\
\hline
280 & 1.13 &361 &1.46 & 368  & 1.49 \\
\hline
310 & 1.26 &398  &1.61 & 402 & 1.63 \\
\hline
340 & 1.38 &434  & 1.76 & 437 & 1.77 \\
\hline
370 & 1.50 &471  & 1.90 & 472 & 1.91 \\
\hline
400 &1.62  & 508  & 2.06 & 507 & 2.05  \\
\hline
\end{tabular}
\caption{Translation table between different temperature identification schemes.
The first two columns display temperatures in the direct scheme, (in which the
 temperature of the holographic model matches the physical QGP temperature)
and the corresponding ratio to the IHQCD critical temperature,  that was
fixed by YM lattice results at $T_c = 247$ MeV \cite{kirfit}.
The third and fourth columns  display the corresponding temperatures
 (and respective ratios to $T_c$)  in the energy scheme, and the last two in the  entropy scheme.
}\label{kirtranslation}
\end{table}

\begin{figure}[h!]
\centering
\includegraphics[scale=1.2]{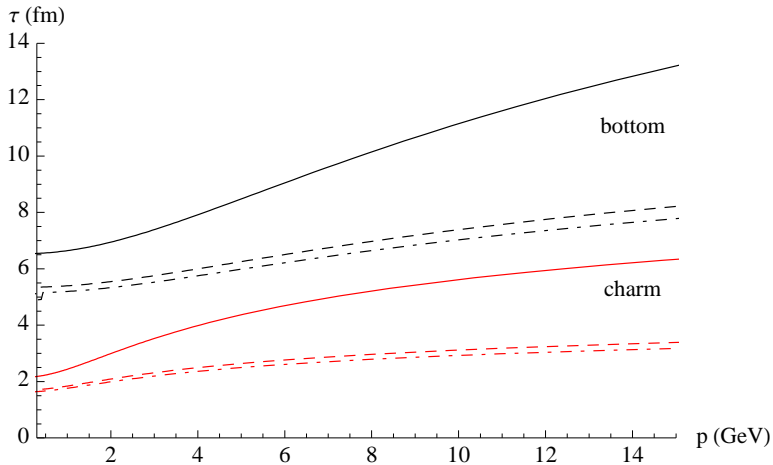}
\caption{Diffusion times for the Charm and Bottom quarks, as a function of
initial momentum, at $T_{QGP}=250 $ MeV. The different lines  represent the
 in the {\em direct} scheme (solid), {\em energy} scheme (dashed)
 and {\em entropy} scheme (dash-dotted), all corresponding to the same temperature $T_{QGP}=250 $ MeV.}\label{kirstopping2}
\end{figure}

\begin{table}[h!]
\centering
\begin{tabular}{|c|c c c|}
\hline
$T_{QGP},MeV$ & $\tau_{charm}$ (fm/c)  & $\tau_{charm}$ (fm/c)  &  $\tau_{charm}$   (fm/c )  \\
& (direct) & (energy) & (entropy) \\
\hline
220  & -  & 4.0  & 3.6   \\
\hline
250 & 5.7 &  3.1 & 3.0 \\
\hline
280 & 4.3   & 2.6  & 2.5  \\
\hline
310 & 3.5 &  2.1 &  2.1  \\
\hline
340 & 2.9 & 1.8 & 1.8 \\
\hline
370 & 2.5 & 1.5   &1.5 \\
\hline
400 & 2.1 &  1.3  & 1.3  \\
\hline
\end{tabular}
\caption{The diffusion times for the charm quark are shown for
different temperatures, in the three different schemes.  Diffusion
times have been evaluated with a quark initial momentum  fixed  at
$p \approx 10$ $GeV$. }\label{kirDiffTablecharm}
\end{table}

\begin{table}[h!]
\centering
\begin{tabular}{|c|c c c|}
\hline
$T_{QGP} (MeV)$ & $\tau_{bottom}$ (fm/c)  & $\tau_{bottom}$ (fm/c)&  $\tau_{bottom}$ (fm/c)  \\
& (direct) & (energy)& (entropy)\\
\hline
220  & -  & 8.9 & 8.4 \\
\hline
250 & 11.4  & 7.5  & 7.1 \\
\hline
280 &  10.1 &  6.3 & 6.1\\
\hline
310 & 8.6  & 5.4  & 5.3 \\
\hline
340 & 7.5 & 4.7 & 4.7  \\
\hline
370 & 6.6 & 4.1  & 4.1  \\
\hline
400 & 5.8 & 3.6 & 3.6\\
\hline
\end{tabular}
\caption{Diffusion times for the  bottom quark are shown for different temperatures, in the three different schemes.
  Diffusion times have been evaluated with a quark initial momentum  fixed  at $p \approx 10$ $GeV$. }\label{kirDiffTablebottom}
\end{table}

The temperature translation table between  the various schemes is shown
in Table \ref{kirtranslation}. In that  table, $T_c = 247 MeV $ is the  deconfining  temperature of the holographic model.

In Figure \ref{kirstopping2} we show the
comparison between the diffusion times, as a function of initial  quark momentum,  in the different  schemes for
the Charm and Bottom quarks, at the  temperature $T_{QGP} = 250 MeV$.

 The results  for the diffusion times at different temperatures,
  computed at  a reference heavy quark initial momentum
$p \approx 10$  $GeV$,  are displayed in Tables \ref{kirDiffTablecharm}
and \ref{kirDiffTablebottom}. We see that there is little practical
difference between the {\em entropy}  and {\em energy} schemes; on the other
hand  the difference between the {\em direct} scheme and the two alternative schemes
can be quite substantial.

\section{Jet quenching parameter}\label{kirjq}
\index{jet quenching}

In this Section we discuss  the jet quenching parameter in the  class of holographic models under consideration, and we estimate its numerical
value for the concrete model with potential (\ref{kirpotential}) and parameters
fixed as in \cite{kirfit}. For the holographic computation, we will  follow \cite{kirLiu1,kirLiu2}.
 There is another method available  \cite{kirGubser-lan}, but we will not use it here.

The jet-quenching parameter $\hat{q}$ provides a measure of the
dissipation of the plasma and it has been associated
 to the behavior of a Wilson loop\index{Wilson loop}  joining two light-like lines.
 We consider two light-like lines  which extend for a distance $L^{-}$ and are situated distance $L$ apart in a transverse coordinate.
Then $\hat{q}$ is given by the large $L^{+}$ behavior of the Wilson loop
\be\label{kirQHATDEF}
W \sim e^{-\frac{1}{4\sqrt{2}}\hat{q}L^{-}L^{2}}\;.
\ee
We consider the bulk string frame metric
\be\label{kirBULKMETRIC}
ds^2 = e^{2A_s(r)}\left(-f(r)dt^{2} +d\vec{x}^{2} +\frac{dr^{2}}{f(r)} \right)  \;.
\ee
To address the problem of the Wilson loop we make a change of coordinates to light cone coordinates for the boundary theory
\be\label{kirLIGHTCONE}
x^{+}=x_1+t\quad x^{-}= x_{1}-t
\ee
for which  the metric becomes
\be\label{kirLIGHTMETRIC}
ds^2 = e^{2A_s}\left(dx_{2}^{2}+dx_{3}^{2}+\frac{1}{2}(1-f)(dx_{+}^{2}+dx_{-}^2)  +(1+f)dx_{+}dx_{-}+\frac{dr^2}{f}\right)\;.
\ee
The Wilson loop in question stretches across $x_{2}$, and lies at a constant $x_{+}$,$x_{3}$.  It is convenient to choose a world-sheet gauge in which
\be\label{kirWORLDGAUGE}
x_{-} =\tau,\quad x_{2} =\sigma\;.
\ee
Then the action of the string stretching between the two lines is given by
\be\label{kirWILSACTION}
S =\frac{1}{2\pi \ls^2} \int d\sigma d\tau \sqrt{-det(g_{MN}\partial_{\alpha}X^{M}\partial_{\beta}X^{N})}
\ee
and assuming a profile of $r=r(\sigma)$ we obtain
\be\label{kirWILSANS}
S = \frac{L^{-}}{2\pi \ls^2} \int d x_{2} ~e^{2A_s}\sqrt{\frac{(1-f)}{2}\left(1+\frac{r'^2}{f}\right)} \;.
\ee
The integrand does not depend explicitly on $x_{2}$, so there is a conserved quantity, $c$:
\be\label{kirCONSQUANT}
r'\frac{\partial\mathcal{L}}{\partial r'} -\mathcal{L} ={c\over \sqrt{2}}
\ee
which leads to
\be\label{kirrp}
r'^2 = f\left(\frac{e^{4A_s}(1-f)}{c^2}-1 \right)\;.
\ee
\begin{figure}[h]
\centering
\includegraphics[width=10cm]{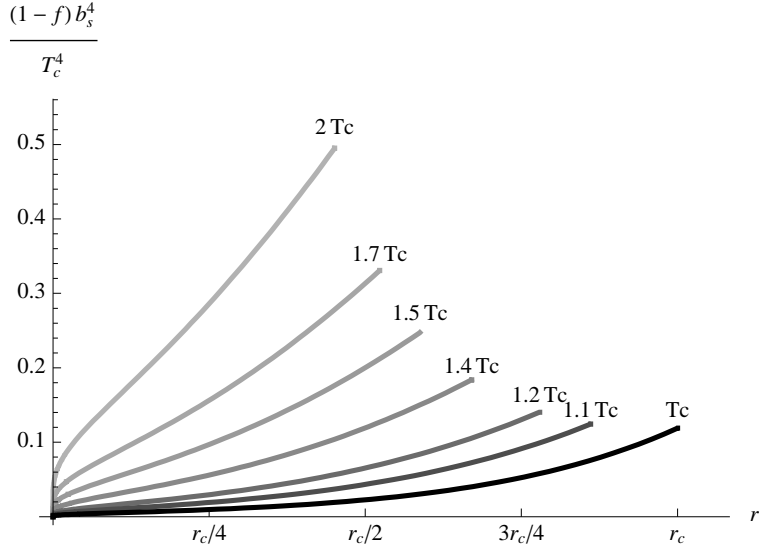}
\caption{In this figure the combination $(1-f)e^{4A_s}$ is plotted
as a function of the radial distance, for several temperatures.
The radial distance is given in units of the horizon position
$r_c$ for the black hole at the critical temperature $T_c$. All
curves stop at the corresponding horizon position.}\label{kirjqpic}
\end{figure}
A first assessment of this relation involves determining the zeros
and the region of positivity of the right-hand side. $f$ is always
positive and vanishes at the horizon. For the second factor we
need the asymptotics of $e^{4A_s}(1-f)$. This factor remains
positive and bounded from below in the interior and up to the
horizon. It vanishes however logarithmically near the boundary as
\be e^{4A_s}(1-f)= \pi T\ell{e^{3A(r_h)}}\left(-{1\over
\beta_0\log(\Lambda r)}\right)^{8\over 3} \left[1+{\cal
O}\left({1\over \log(\Lambda r)}\right)\right] \label{kirb15}\ee This
is unlike the conformal case where we obtain a constant \be
e^{4A_s}(1-f)\Big |_{\rm conformal}=(\pi T\ell)^4 \label{kirb16}\ee

Because of this, for fixed $c$, there is a region near the
boundary where $r'^2$ becomes negative. At this stage we will
avoid this region, by using a modified boundary at $r=\epsilon$.
We will later show that this gymnastics will be irrelevant for the
computation of the jet quenching parameter, as it involves
effectively the limit $c\to 0$.

We will place the modified boundary $r=\epsilon$ a bit inward from
the place $r=r_{min}$ where the factor
$\frac{e^{4A_s}(1-f)}{c^2}-1$ vanishes: \be
e^{4A_s(r_{min})}(1-f(r_{min}))=c^2 \label{kirb17}\ee Therefore we
choose $r_{min}<\epsilon$.

Then, in the range $\epsilon <r<r_h$ the factor $\frac{e^{4A_s}(1-f)}{c^2}-1$ is positive for sufficiently  small $c$.
In this same range, $r'$ vanishes only at  $r=r_h$.
This is the  true turning point of the string world-sheet.
 This is also what happens in the conformal case.
 It is also intuitively obvious that the relevant Wilson loop must sample
also the region near the horizon.

The constant $c$ is determined by the fact that the two light-like Wilson loops are a $x_{2}=L$ distance apart.
\be
\frac{L}{2} =\int_{\epsilon}^{r_{h}} \frac{c dr}{\sqrt{f(e^{4A_s}(1-f)-c^2)}}\;.
\label{kirb18}\ee
The denominator vanishes at the turning point. The singularity is
  integrable\footnote{Even if we choose $\epsilon=r_{min}$, the new singularity at $r=r_{min}$ is also integrable as suggested from
(\ref{kirb15}).}.
Therefore, as we are interested in the small $L$ region, it is obvious from the expression above that that
$c$ must also be small in the same limit.

This relation can then be expanded in powers of $c$ as
\be
\frac{L}{2c} =\int_{\epsilon}^{r_{h}} \frac{e^{-2A_s}dr}{\sqrt{f(1-f)}}+{c^2\over 2}
\int_{\epsilon}^{r_{h}} \frac{e^{-6A_s}dr}{\sqrt{f(1-f)^3}}+{\cal O}(c^4)\;.
\label{kirb19}\ee
Therefore to leading order in $L$
\be
c={L\over 2\int_{\epsilon}^{r_{h}} \frac{e^{-2A_s}dr}{\sqrt{f(1-f)}}}+{\cal O}(L^3)
\label{kirb20}\ee

We are now ready to evaluate the Nambu-Goto\index{Nambu-Goto action} action of the extremal configuration we have found.
Starting from (\ref{kirWILSANS}), we substitute $r'$ from  (\ref{kirrp}), and change integration variable from $x_2 \to r$  to obtain
\be
S = \frac{2L^{-}}{2\pi \ls^2} \int_{\epsilon}^{r_h} dr ~{e^{4A_s}(1-f)\over \sqrt{2f\left(e^{4A_s}(1-f)-c^2\right)}} \;.
\label{kirb23}\ee

 As in \cite{kirLiu1,kirLiu2}, we
subtract from equation (\ref{kirb23}) the action of two free string straight world-sheets that
hang down to the horizon. To compute this action a convenient
choice of gauge is $x_{-} =\tau,\quad r =\sigma$. The action of
each sheet is \be S_{0}={L^-\over 2\pi
\ls^2}\int_{\epsilon}^{r_h}~dr~\sqrt{g_{--}g_{rr}}= {L^-\over 2\pi
\ls^2}\int_{\epsilon}^{r_h}~dr~e^{2A_s}\sqrt{1-f\over 2f}
\label{kirb24}\ee

The subtracted  action is  therefore:

\be\label{kirACTIONSMALLC}
S_{r}= S-2S_{0} = \frac{L^{-}c^2}{2\pi \ls^2} \int_{\epsilon}^{r_h} \frac{dr}{e^{2A_{s}}\sqrt{f(1-f)}}+{\cal O}(c^4)\;,
\ee
Using now (\ref{kirb20}) to substitute $c$ we finally obtain
\be\label{kirACTIONSMALLL}
S_{r} =\frac{L^{-} L^{2}}{8 \pi \ls^2} \frac{1}{\int_{\epsilon}^{r_h} \frac{dr}{e^{2A_{s}}\sqrt{f(1-f)}}}+{\cal O}(L^4)\;.
\ee

So far we have evaluated the relevant Wilson loop\index{Wilson loop} in the
fundamental representation (by using probe quarks). On the other
hand, the Wilson loop that defines  the jet-quenching parameter is
an adjoint one. We can obtain it in the large-$N_c$ limit from the
fundamental using $tr_{\rm Adjoint} =tr_{\rm Fundamental}^2$. We
finally extract the jet-quenching parameter\index{jet quenching}  as \be\label{kirWILSON}
\hat{q} = \frac{\sqrt{2}}{\pi \ls^2}\frac{1}{\int_{\epsilon}^{r_h}
\frac{dr}{e^{2A_s}\sqrt{f(1-f)}}}\;. \ee We are now ready to
remove the cutoff. As the integral appearing is now well-defined
up to the real boundary $r=0$ we may rewrite it as \be
\int_{\epsilon}^{r_{h}}
\frac{e^{-2A_s}dr}{\sqrt{f(1-f)}}=\int_{0}^{r_{h}}
\frac{e^{-2A_s}dr}{\sqrt{f(1-f)}}-I(\epsilon) \sp
I(\epsilon)=\int_0^{\epsilon} \frac{e^{-2A_s}dr}{\sqrt{f(1-f)}}
\label{kirb21}\ee

In \cite{kirdrag} we obtain the small
$\epsilon$ estimate of $I(\epsilon)$ that vanishes as $\sim
\epsilon (\log\epsilon)^{4\over 3}$. We may finally
write\footnote{In practise, the previous discussion including
regularizing the UV is academic. The numerical calculation is done
with a finite cutoff where the boundary conditions for the
couplings are imposed.} \be\label{kirb22} \hat{q} =
\frac{\sqrt{2}}{\pi \ls^2}\frac{1}{\int_{0}^{r_h}
\frac{dr}{e^{2A_s}\sqrt{f(1-f)}}}\;. \ee

\begin{figure}[h!]
\centering
\includegraphics[width=9cm]{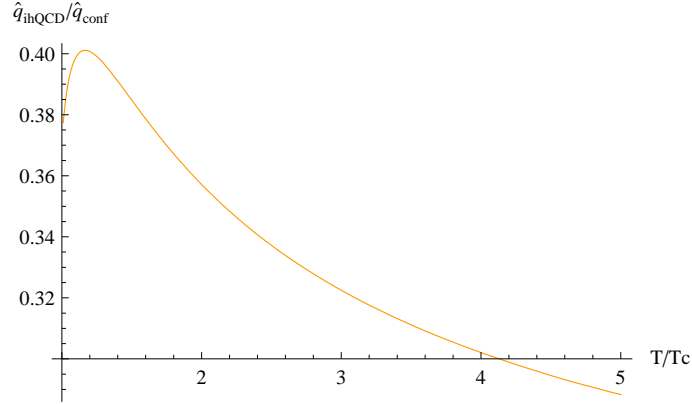}
\caption[]{In this figure the ratio of the jet quenching parameter
in our model to the jet quenching parameter in ${\cal N}=4$ is shown.
The integral present in equation (\ref{kirWILSON}) has been numerically calculated from
an effective cutoff at $r=r_{h}/1000$.  The jet quenching parameter in
${\mathcal N}=4$ SYM has been calculated with $\l_{'t Hooft}=5.5$.
}\label{kirMyFigureQhatA}
\end{figure}

\begin{figure}[h!]
\centering
\includegraphics[width=9cm]{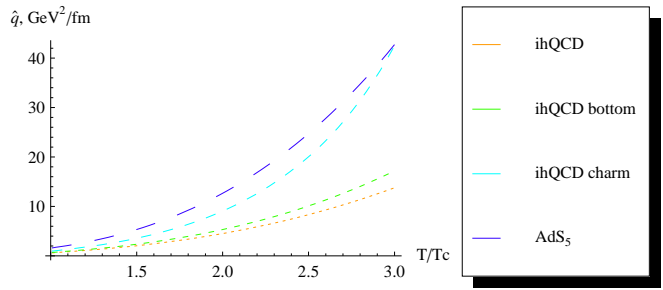}
\caption{ The jet quenching parameter $\hat{q}$ for the Improved
Holographic QCD model and $\mathcal{N}=4$ SYM is shown in units of
$GeV^2/fm$ for a region close to $T=T_{c}$.  The smallest dashed
curve is the ihQCD result with an effective cutoff of $r_{cutoff}=r_{h}/1000$.
The small dashed curve is the ihQCD result with the cutoff from the mass of the Bottom quark.
 The medium dashed curve has a cutoff coming from the Charm mass and
 and largest dashed curve is the $\mathcal{N}=4$ SYM result.}\label{kirMyFigureQhatBTwo}
\end{figure}

\begin{figure}[h!]
\centering
\includegraphics[width=9cm]{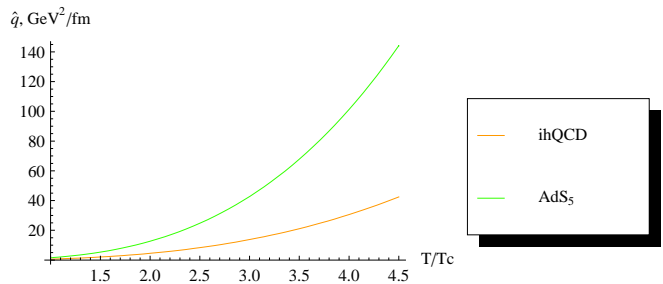}
\caption{The jet quenching parameter $\hat{q}$ for the Improved
 Holographic QCD model (lower curve) and $\mathcal{N}=4$ SYM (upper curve)
  are shown in units of $GeV^2/fm$ for temperatures up to $T=4T_c$.}\label{kirMyFigureQhatB}
\end{figure}

\begin{table}[h!]
\centering
\begin{tabular}{|c|c c |}
\hline
$T_{QGP},MeV$ & $\hat{q}$ $(GeV^2/fm)$  & $\hat{q}_{1}$ $(GeV^2/fm)$  \\
& (direct) & (direct)  \\
\hline
220  & -  & -    \\
\hline
250 & 0.5 & 0.6   \\
\hline
280 & 0.8 & 0.8   \\
\hline
310 & 1.1 & 1.1    \\
\hline
340 & 1.4 & 1.4  \\
\hline
370 & 1.8 & 1.8 \\
\hline
400 & 2.2 & 2.2  \\
\hline
\end{tabular}
\caption{
This table shows the jet quenching parameter $\hat{q}$ computed
 with different cutoffs for the different temperatures shown in the first column.
 The computation is done in the direct scheme.  The second column shows $\hat{q}$
 with a cutoff at $r_{cutoff}=r_{h}/1000$, where $r_{h}$ is the location
  of the horizon. In accordance with the conclusions of appendix
 $\hat{q}$ does not change significantly as we vary the cutoff from
 $r_{h}/1000$ to $r_{h}/100$.
}\label{kirQHATCUTOFF}
\end{table}

\begin{table}[h!]
\centering
\begin{tabular}{|c|c c c|}
\hline
$T_{QGP},MeV$ & $\hat{q}$ $(GeV^2/fm)$  & $\hat{q}$ $(GeV^2/fm)$  &  $\hat{q}$ $(GeV^2/fm)$   \\
& (direct) & (energy) & (entropy) \\
\hline
220  & -  & 0.9  & 1.0   \\
\hline
250 & 0.5 & 1.2  & 1.3 \\
\hline
280 & 0.8 & 1.6  & 1.7  \\
\hline
310 & 1.1 & 2.1  & 2.2   \\
\hline
340 & 1.4 & 2.7 & 2.8 \\
\hline
370 & 1.8 & 3.4 & 3.4 \\
\hline
400 & 2.2 & 4.2 & 4.2 \\
\hline
\end{tabular}
\caption{This table displays the jet quenching parameter $\hat{q}$ using the three
 different comparison schemes. For lower temperatures the ``entropy scheme`` gives
  higher values.  As energy is increased the energy and entropy schemes temperatures start to
 coincide and there is little difference in the jet quenching
 parameter as well.}\label{kirQHATSCHEMES}
\end{table}

\begin{table}[h!]
\centering
\begin{tabular}{|c|c c c|}
\hline
$T_{QGP},MeV$ & $\hat{q}_{charm}$ $(GeV^2/fm)$  & $\hat{q}_{charm}$ $(GeV^2/fm)$  &  $\hat{q}_{charm}$ $(GeV^2/fm)$   \\
& (direct) & (energy) & (entropy) \\
\hline
220  & -  & 1.3   & 1.5 \\
\hline
250 & 0.8 & 1.8  & 2.0  \\
\hline
280 & 1.2 & 2.6  & 2.8 \\
\hline
310 & 1.7 & 3.5  & 3.6 \\
\hline
340 & 2.2 & 4.6 & 4.7 \\
\hline
370 & 2.8 & 5.9 & 6.0 \\
\hline
400 & 3.6 & 7.6 & 7.5 \\
\hline
\end{tabular}
\caption{This table displays the jet quenching parameter $\hat{q}$ using the three different comparison schemes with an effective cutoff provided by the mass of the Charm quark. Again, for lower temperatures the ``entropy scheme`` gives higher values.  As energy is increased the energy and entropy schemes temperatures start to coincide and there is little difference in the jet quenching parameter as well.
  Also when the temperature approaches the quark mass the
   picture of the heavy quark as a hanging string collapses and results are not reliable.}\label{kirQHATCHARM}
\end{table}

\begin{table}[h!]
\centering
\begin{tabular}{|c|c c c|}
\hline
$T_{QGP},MeV$ & $\hat{q}_{bottom}$ $(GeV^2/fm)$  & $\hat{q}_{bottom}$ $(GeV^2/fm)$  &  $\hat{q}_{bottom}$ $(GeV^2/fm)$   \\
& (direct) & (energy) & (entropy) \\
\hline
220  & -  & 1.0   & 1.1  \\
\hline
250 & 0.6 & 1.4  & 1.5  \\
\hline
280 & 0.9 & 1.9  & 2.0 \\
\hline
310 & 1.2 & 2.5  & 2.6 \\
\hline
340 & 1.6 & 3.2 & 3.2 \\
\hline
370 & 2.0 & 4.0  & 4.0 \\
\hline
400 & 2.5 & 5.0  & 4.9 \\
\hline
\end{tabular}
\caption[]{
 This table displays the jet quenching parameter $\hat{q}$ using the three different comparison schemes with an effective cutoff
  provided by the mass of the Bottom quark. The results are close to the $\hat{q}$ results computed in Table  \ref{kirQHATSCHEMES}
since the mass of the Bottom quark is much larger than the temperatures we examine.
}\label{kirQHATBOTTOM}
\end{table}
\index{jet quenching}

From equation (\ref{kirb22}) we obtain, in the conformal case:
\be
\hat q_{\rm conformal}={\Gamma\left[{3\over 4}\right]\over \Gamma\left[{5\over 4}\right]}~\sqrt{2\l}~\pi^{3\over 2}T^3
\ee
The conformal value, for the median value of $\lambda=5.5$ and $T\simeq 250$ MeV
 gives $\hat q_{\rm conformal}\simeq 1.95$ GeV$^2$/fm where we used the conversion
$1$ GeV$\simeq 5$ fm$^{-1}$.

Numerical evaluation of equation (\ref{kirb22}) in the non-conformal IHQCD setup\footnote{In
this case, the value of $\ell_s$ appearing in equation (\ref{kirb22}) is fixed as explained in Section 4.}
gives us a  value of $\hat{q}$ which is lower (at a given temperature) than the
 conformal value, as shown in Figures \ref{kirMyFigureQhatA},
\ref{kirMyFigureQhatBTwo} and \ref{kirMyFigureQhatB}. Tables \ref{kirQHATCUTOFF}
to \ref{kirQHATBOTTOM} display the numerical values of the jet quenching
parameter at different temperatures in the experimentally relevant range,
in different temperature matching schemes.

\section{Discussion and outlook}\label{kirconclusions}

The construction presented in this paper offers a holographic description of large-$N_c$ Yang-Mills theory that
is both realistic and calculable, and in quite good agreement with a large number of lattice results
both at zero and finite temperature.

It is a phenomenological model; as such it is not directly associated to an
explicit string theory construction.
In this respect it is  in the same class as the models
based  on pure AdS backgrounds (with hard or soft walls) \cite{kiradsqcd1,kirsoft,kirherzog,kirbraga},
 or on IR deformations of the AdS metric \cite{kiradr,kirkaj}.
In comparison to them however, the IHQCD approach  has the
advantage that the dynamics responsible for strong coupling phenomena (such as
confinement and phase transitions)  is made  explicit in the bulk description,
and  it is tied  to  the fact that the coupling constant depends on the energy scale and becomes
large in the IR. This makes the model consistent and calculable,
once the 5D effective action is specified: the dynamics can be entirely derived from the bulk Einstein's equation.
The emergence of an IR mass scale and the finite temperature phase structure are built-in:
they  need not be imposed by hand and
do not suffer from ambiguities related to IR boundary conditions (as in hard wall models) or from inconsistencies
in the laws of thermodynamics (as in non-dynamical soft wall models
based on a fixed dilaton profile \cite{kirsoft} or on a fixed metric \cite{kiradr,kirkaj}).
More specifically,  in this approach it is guaranteed that
the Bekenstein-Hawking temperature of the black hole\index{black hole} matches the entropy computed as the derivative of the free energy
with respect to temperature.

With an appropriate choice of the potential, a realistic and quantitatively accurate description of
essentially all the static properties (spectrum and equilibrium thermodynamics) of the dynamics of pure Yang-Mills can be provided.
The main ingredient responsible for the
dynamics (the dilaton potential) is fixed through comparison  with both perturbative QCD and
lattice results. It is worth stressing that  such a matching on the quantitative level was  only possible
because the class of holographic models we discuss {\em generically} provides a {\em qualitatively accurate}
description of the strong Yang-Mills dynamics. This is a highly  non-trivial fact, that strongly indicates
that a realistic  holographic description of real-world QCD might be ultimately possible.

Although the asymptotics of our potential is dictated by general principles, we base our choice of parameters
by comparing with the lattice results for the thermodynamics.
There are other physical parameters in the 5D description  that do not appear in the potential:
the 5D Planck scale, that was fixed by matching the free field thermodynamics
in the limit  $T\to \infty$; the coefficients in the axion kinetic term, that were set by matching
the axial glueball spectrum and the topological susceptibility (from the lattice).  The quantities that
we use as input in our fit, as well as the corresponding parameters in the 5D model, are shown
in the upper half of Table \ref{kirresults}.

The fact that our  potential has effectively two free parameters depends on our choice of the functional form. This
functional form contains some degree of arbitrariness, in that only
 the UV and IR asymptotics of $V(\l)$  are fixed by general considerations
(matching the perturbative $\beta$-function in the UV, and  a discrete linear glueball spectrum for the IR).
Therefore the results presented in this paper offer more a {\em description}, rather than a {\em prediction} of the
thermodynamics.

Nevertheless, there are several quantities that we successfully ``postdict'' (i.e. they agree with
the lattice results) once the potential is fixed: apart from the good agreement of the thermodynamic functions
over the whole range of temperature explored by the lattice studies (see Figures \ref{kiresp}, \ref{kirtrace2} and \ref{kirsound}),
they are the lowest glueball mass ratios and the value
of the critical temperature.  The comparison  of these quantities with the lattice results is shown in the lower half
of Table \ref{kirresults}, and one can see that the agreement is overall very good.
Moreover the model predicts the masses of the full towers of glueball states in the $0^{+-},0^{++},2^{++}$ families.

The fact that IHQCD is consistent with a large number of lattice results is clearly
not the end of the story: its added value, and one of the main reasons for its interest lies in
its immediate applicability beyond equilibrium thermodynamics, i.e. in the dynamic regimes tested
in heavy-ion collision experiments. This is a generic feature of the holographic approach, in which
there are no obstructions (as opposed to the lattice) to perform real-time computations and to calculate
hydrodynamics and transport coefficients.
IHQCD provides a framework to compute these quantities
in a case where the static properties agree with the real-word  QCD
at the quantitative level. Therefore the bulk viscosity, drag force and jet quenching parameters were computed in IHQCD:

\paragraph{Bulk Viscosity:} The bulk viscosity was computed by calculating the low frequency
 asymptotics of the appropriate stress tensor correlator
holographically. The result is that the bulk viscosity rises near the phase transition but stays always below the shear viscosity.
It floats somewhat above the Buchel bound, with a coefficient of proportionality varying between 1 and 2.
Therefore it is expected to affect the elliptic flow at the small percentage level \cite{kirHeinz,kirheinz}.
Knowledge of the bulk viscosity is important in extracting the shear viscosity from the data.
This result is not in agreement with the lattice result near $T_c$. In particular
the lattice result gives a value for the viscosity that is
ten times larger.

The bulk viscosity\index{viscosity!bulk} keeps increasing in the black-hole branch below
the transition point until the large BH turns into the small BH at
a temperature $T_{min}$.  The bulk viscosity on the small BH
background is always larger than the respective one in the large
BH background.  In particular, it can be shown
that the T derivative of the quantity $\zeta/s$ diverges at
$T_{min}$. This is the holographic reason for the presence of a peak
in $\zeta/s$ near $T_c$. On the other hand, as it is shown in
\cite{kirGKMN2}, the presence of $T_{min}$ (i.e. a small BH branch) is in
one-to-one correspondence with color confinement at zero T.
We arrive thus at the suggestion that in
a (large N) gauge theory that confines color at zero T, there shall
be a rise in $\zeta/s$ near $T_c$.

An important ingredient here is the value of the viscosity  asymptotically in the small BH branch. There
the asymptotic value correlates to the IR behavior of the potential. Taking also into account the fact that this asymptotic
value is very close to the value of the bulk viscosity near $T_c$,
 we can derive bounds that suggest that the bulk viscosity cannot increase
a lot near $T_c$.

\paragraph{Drag Force:} The drag force  calculated from IHQCD has the expected behavior.
 Although it increases with temperature, it does so slower than in ${\cal N}=4$ SYM, signaling the
effects of asymptotic freedom.

\paragraph{Diffusion Time:}Based on the drag force calculation the diffusion times can be computed for a heavy external quark.
  The numerical values obtained are in agreement with phenomenological models \cite{kirlan}.
   To accommodate for the fact that IHQCD exhibits a phase transition around $T=247$ $MeV$
   (i.e. about $30\%$ higher than in QCD), the  results are compared using alternative schemes, as  proposed in \cite{kirgub}.
For example, for an external Charm quark of momentum $p=10$ $GeV$  (in
the alternative scheme) a diffusion time of $\tau=2.6$ $fm$ at temperature $T=280$ $MeV$ is found.
  Similarly, for a Bottom quark of the same momentum and at the same temperature, $\tau=6.3$ $fm$.
  Generally the numbers obtained are close to those obtained by \cite{kirlan} and \cite{kirlan2}.

\paragraph{Jet Quenching:}The jet quenching parameter of this model, has been also calculated,
based on the formalism of \cite{kirLiu1,kirLiu2} by computing the appropriate light-like Wilson loop.
 $\hat{q}$ grows with temperature, but slower than the $T^3$ growth of $\mathcal{N}=4$ SYM result.
Again this can be attributed to the incorporation of asymptotic freedom in IHQCD.
 Using the alternative scheme to compare with experiment, the results
 are close to the lower quoted values of $\hat{q}$.
   For example, for a temperature of $T=290$ MeV, which in the alternative ``energy scheme'' corresponds to a temperature of
   $T=395$ MeV in our model, we find that $\hat{q}\approx 2 GeV^2/fm$.

   However, the numbers obtained for this particular definition of jet quenching parameter seem rather low and indicate
   that this may not be the most appropriate definition in the holographic context.
   There are other ways to define $\hat q$, in particular using the fluctuations of the trailing string solution.
   This is gives a direct and more detailed input in the associated Langevin dynamics and captures the asymmetry
    between longitudinal and transverse fluctuations.
    This can be computed along the lines set in 
    \cite{kirGubser-lan,kirson,kiriancu} and the results were recently reported in \cite{lange}.

\newpage
 \addcontentsline{toc}{section}{Acknowledgements}
\vskip .2cm
\centerline{\bf Acknowledgments}
\vskip 1cm

We would like to thank the numerous colleagues that have shared their insights with us via discussions and correspondence
since 2007 that this line of research has started:
O. Aharony, L. Alvarez-Gaum\'e,  B. Bringoltz, R. Brower, M. Cacciari, J. Casalderrey-Solana, R. Emparan , F. Ferrari, B. Fiol,  P. de Forcrand,
 R. Granier de Cassagnac, L. Giusti, S. Gubser,   K. Hashimoto, T. Hertog, U. Heinz, D. K. Hong, G. Horowitz,  E. Iancu,
  K. Intriligator, K. Kajantie, F. Karsch, D. Kharzeev, D.  Kutasov, H. Liu, B. Lucini,
  M. Luscher,  J. Mas, D. Mateos, H.B. Meyer,  C. Morningstar, V. Niarchos,
C. Nunez, Y. Oz, H. Panagopoulos,  S. Pal, M. Panero,
 I. Papadimitriou,   A. Paredes,   A. Parnachev, G. Policastro, S. Pufu, K. Rajagopal, F.Rocha,
 P. Romatchke, C. Salgado,  F. Sannino, M. Shifman, E. Shuryak, S. J. Sin,
 C. Skenderis, D. T. Son,  J. Sonnenschein, S. Sugimoto,  M. Taylor, M. Teper, J. Troost,
 A. Tseytlin, A. Vainshtein,
G. Veneziano, A. Vladikas,  L. Yaffe, A. Yarom   and U. Wiedemann.

This work was partially supported by a European Union grant FP7-REGPOT-2008-1-CreteHEPCosmo-228644,
and ANR grant NT05-1-41861.
Work of LB has been partly funded by INFN, Ecole Polytechnique (UMR du CNRS 7644),
MEC and FEDER under grant FPA2008-01838, by the Spanish Consolider-Ingenio 2010 Programme CPAN
(CSD2007-00042) and by Xunta de Galicia (Conseller\'ia de Educaci\'on and grant\\ PGIDIT06PXIB206185PR).

E. K. thanks the organizers and especially E. Papantonopoulos for organizing a very interesting and stimulating school.

\end{document}